\documentclass[final,3p]{elsarticle}

\usepackage{lineno,hyperref}
\usepackage{amsmath}
\usepackage{amsfonts}
\usepackage{amssymb}
\usepackage{algorithm}
\usepackage{algorithmic}
\usepackage{subfigure}
\usepackage{caption}
\usepackage{graphicx}
\usepackage{epstopdf}
\usepackage{epsfig}
\usepackage{threeparttable}
\usepackage{multirow}

\journal{Expert Systems with Applications}

\newcommand{\norm}[1]{\left\Vert#1\right\Vert}

\newcommand{\pref}[1]{(\ref{#1})}
\begin{document}

\begin{frontmatter}

\title{Manifold Feature Index: A novel index based on high-dimensional data simplification}


\author[add1]{Chenkai Xu}
\ead{ckxu@zju.edu.cn}
\author[add1,add2]{Hongwei Lin}
\ead{hwlin@zju.edu.cn}
\author[add1]{Xuansu Fang}
\ead{793137669@qq.com}

\address[add1]{School of Mathematics Science, Zhejiang University, Hangzhou, 310027, China}
\address[add2]{State Key Laboratory of CAD$\&$CG, Zhejiang University, Hangzhou, 310058, China}

\begin{abstract}

In this paper,
  we propose a novel stock index model,
  namely the manifold feature(MF) index,
  to reflect the overall price activity of the entire stock market.
Based on the theory of manifold learning,
  the researched stock dataset is assumed to be a low-dimensional manifold embedded in a higher-dimensional Euclidean space.
After data preprocessing,
  its manifold structure and discrete Laplace-Beltrami operator(LBO) matrix are constructed.
We propose a high-dimensional data feature detection method to detect feature points on the eigenvectors of LBO,
  and the stocks corresponding to these feature points are considered as the constituent stocks of the MF index.
Finally,
  the MF index is generated by a weighted formula using the price and market capitalization of these constituents.
The stock market studied in this research is the Shanghai Stock Exchange(SSE).
We propose four metrics to compare the MF index series and the SSE index series
  (SSE 50, SSE 100, SSE 150, SSE 180 and SSE 380).
From the perspective of data approximation,
    the results demonstrate that our indexes are closer to the stock market than the SSE index series.
From the perspective of risk premium,
   MF indexes have higher stability and lower risk.
\end{abstract}


\end{frontmatter}


\section{Introduction}


A stock market index or stock index is a statistic that reflects the overall price activity of the stock market.
It is typically designed by stock exchanges or financial services
  and computed by the prices of selected stocks (constituents) with weights.
For investors and financial managers,
  a stock market index is a tool to design portfolios,
  to test investment results,
  and to forecast market trends.
For companies, news agencies, and government agencies,
  it is a tool to observe and predict the development trends of social politics and economics.
Furthermore,
  it is the basis of some financial derivatives,
  such as stock index futures and stock index options,
  which provide more diverse investment strategies and reduce the risk of investment.

A primary criterion of the stock market index is transparency~\cite{lo2016index}.
For this reason,
  the two main parts of computing a stock market index,
  selecting constituent stocks and presetting the weights
  must be published by the stock exchanges or financial service providers who design the stock market index.
Some stock indexes such as the Shanghai Stock Exchange composite index (SSE),
  whose constituents include all the stocks in the Shanghai Stock Exchange,
  do not need to select the constituent stocks.
However most existing stock indexes such as
  the Dow Jones Industrial Average (DJIA) index and
  the National Association of Securities Dealers Automated Quotations (NASDAQ) composite index
  select constituents from a larger stock sample space.
The selection criteria include
  market capitalization (market cap), free float ratio, sales/earnings (S/E), net profit, and other financial factors.
The constituents are typically determined by experts in exchanges or financial services,
  after ranking them based on the aforementioned criteria.
After selecting the constituents of the stock index,
  the next step is to preset the weights to composite them as an index.
Besides simply using market cap (SSE),
  the SSE index series (SSE 50, SSE 100, SSE 150, etc.) and DJIA add free float to determine the weights.
In addition,
  there are some indexes that consider more weighting factors,
  including cash flow, dividends, book value (Financial Times Stock Exchange index series, FTSE),
  growth variables,
  and value variables (S\&P index series).

For existing stock index models,
   the selection of constituents typically
   considers financial factors such as market cap and free float,
   regardless of stock price changes.
This means that the information on price activity is not used when selecting constituents.
Meanwhile,
  the selection strategy is typically based on ranking
  and finally determined by experts in exchanges or financial services.
Therefore there are subjective factors involved.
In this paper,
  we propose a novel method for selecting constituents,
  that considers more information on price activity,
  and specifies an objective selection rule.
Our index is named as the manifold feature(MF) index
  as we treat stock dataset as a point cloud set embedded in the manifold,
  and the selection method is based on the structure of the manifold itself.

Our study includes all the stocks in the Shanghai Stock Exchange.
After data preprocessing,
  the closing price curves of all stocks are represented as high-dimensional vectors with the same dimensionality.
As per selection rules,
  each stock is regarded as a high-dimensional point,
  and the entire stocks dataset is regarded as a point-cloud in a high-dimensional Euclidean space.
Based on the theory of manifold learning~\cite{balasubramanian2002isomap,donoho2003hessian},
  the high-dimensional dataset
  can be regarded as points in a low-dimensional manifold,
  embedded in a higher-dimensional Euclidean space.
Using the construction method mentioned in ~\cite{belkin2003laplacian},
  we establish the connection between the stocks and construct its discrete Laplace-Beltrami operator (LBO) matrix.
The eigenfunctions of the LBO,
  which can be regarded as scalar fields defined on the manifold,
  present the geometric features of the manifold~\cite{Shen2006Spectral}.
We propose a high-dimensional data feature detection method to select constituents using these eigenvectors.
In our selection method,
  we detect the local maximum and minimum points on the eigenvectors as feature points of the entire dataset.
The detection starts from the eigenvectors with the smallest eigenvalue,
  and stops when the required number of points is reached.
As each feature point corresponds to a stock in the SSE,
  the points (stocks) that we detect are regarded as the selected constituents from the complete stock space.
Finally,
  the MF index is computed using the existing weighting method,
  similar to that of the SSE composite index.



Our contributions are twofold.
   \begin{itemize}
      \item  We propose a new stock index,
               namely the MF index,
               whose method of constituent selection is completely new and different from existing stock indexes.
             The method regards the stocks as points in a low-dimensional manifold
               embedded in a high-dimensional Euclidean space,
               and selects constituents based on the manifold structure itself.
      \item  To select the constituents,
                 we propose a high-dimensional data feature detection algorithm based on the LBO of the manifold.
             The algorithm detects the local maximum and minimum points on eigenvectors as feature points of the entire dataset,
                  and the stocks corresponding to these feature points
                  are regarded as the constituents selected from the complete stock space.
    \end{itemize}

In addition,
  to quantitatively measure the quality of our index,
  this study employs four metrics including
  \emph{Pearson correlation coefficient},
  \emph{Alpha},
  \emph{Beta},
  and \emph{Jensen's alpha}.
They are used to analyze our index in the experiments,
  for the aspects of data approximation and risk premium.

The remainder of this paper is organized as follows.
Section 2 reviews related work on the stock index,
  the applications of LBO
  and the metrics we use to measure the quality of our index.
Section 3 introduces the detailed procedures for generating the MF index
  including data preprocessing,
  the theory and computation of high-dimensional data simplification,
  which is used to determine the constituents of the MF index,
  as well as the weighting formula.
Moreover,
  we propose an algorithm for our method.
In Section 4,
  the four metrics that are used to measure the quality of the MF index are introduced.
Section 5 presents the experimental results, detailed analyses, and discussions.
Section 6 concludes the paper and presents further discussion.

\section{Related work}

In this section,
  we review related work in three aspects.
Firstly,
  we address the works that are related to the stock index model,
  including stock indexes that are mainly used in global stock markets,
  and novel index models proposed in previous papers and patents.
Secondly,
  we review manifold learning on high-dimensional data,
  which is the core idea of our work.
Moreover,
  the applications of LBO on a manifold are introduced.
Finally,
  the studies and applications of the four metrics,
   Pearson correlation coefficient,
   Alpha,
   Beta,
   and Jensen's Alpha
   are reviewed,
  because we use these metrics to compare our index with current indexes,
  in our experiments.

\subsection{Stock index}
Capitalization-weighted index
  is a stock market index wherein the constituents are weighted
  according to the total market capitalization (or free-float market capitalization) of their outstanding shares.
It is widely used in global stock markets
  including DJIA,
  NASDAQ,
  S\&P 500,
  SSE,
  and HSI.
According to Haugen et al.~\cite{haugen1991efficient},
  a capitalization-weighted index is inefficient
  when investors disagree about the risk and expected returns,
  when short selling is restricted,
  when investment income is taxed,
  among other reasons.
There are already a small number of stock indexes weighted by a fundamentally based method,
  which considers a company's economic fundamental factors rather than
  the company's listed market value~\cite{wang2009research}.
Dow Jones Global Titans 50 (DJGT),
Dow Jones Sector Titans Index (DJST),
  and Dow Jones Asian Titans 50 (DJAT) use two fundamental factors as
  criteria for selection,
  sales/earnings and net profit.
The RAFI (Research Affiliates) series in the FTSE index series uses sales/earnings,
  cash flow,
  dividends, and book value,
  which are fundamental factors,
  as the criteria for selection~\cite{granger1969investigating}.

Meanwhile,
  some papers and patents have proposed new stock index models.
Wang et al.~\cite{wang2009research} proposed a fundamental stock price constituent index (JOYFI 300),
  which uses decision trees and logistic regression to select the constituents and compute the weights.
Fernholz et al.~\cite{fernholz1998diversity} proposed a diversity-weighted index,
  which has the desirable characteristics of a passive large-stock index,
  and a performance advantage over a capitalization-weighted index,
   under conditions of a neutral diversity change.
Arnott et al.~\cite{arnott2009non} proposed a non-capitalization weighted index model.
Unlike typical market capitalization weighting and price weighting,
   it uses financial metrics such as book value,
   sales,
   revenue,
   earnings,
   and some non-financial metrics.
Arnott's model avoids overexposure to overvalued securities and underexposure to undervalued securities.
Sauter et al.~\cite{sauter2009method} provided a computer-implemented method to design an index,
  which uses a systematic stock migration process to add or remove stocks from an index.
One advantage of Sauter's model is
  that the number of stocks in the index need not be a fixed value.

\subsection{Manifold learning and applications of LBO}
The basic idea of manifold learning is to assume that the high-dimensional data lie on
  an embedded non-linear manifold in the higher-dimensional space~\cite{balasubramanian2002isomap,donoho2003hessian},
  which is employed in this study to analyze stock data.
A typical application of manifold learning is non-linear dimensionality reduction,
  such as isomap~\cite{balasubramanian2002isomap},
  LLE~\cite{donoho2003hessian},
  and Laplace eigenmaps~\cite{belkin2003laplacian}.
Manifold learning is also widely used in other fields,
 including signal processing,
  image classification,
  energy engineering data analysis,
  medical image data,
  and financial data dimensionality reduction
  ~\cite{wang2015detection,tang2014fault,huang2014kernel,huang2017nonlinear,guerrero2017group,lu2015multi,changpinyo2016synthesized}.
Our study focuses not only on the manifold itself,
  but also on the LBO on the manifold,
  which is helpful for analyzing the structure of the manifold from the perspective of the frequency domain.
For the Fourier analysis of a manifold~\cite{vallet2008spectral},
  the eigenvalues of an LBO specify the discrete frequency domain of a manifold,
  and the eigenfunctions are extensions of the basis functions.
They have been widely used in
   shape analysis~\cite{reuter2006laplace,ovsjanikov2010one},
   shape matching~\cite{sharma2010shape},
   shape retrieval~\cite{reuter2006laplace,bronstein2011shape}
   and mesh processing~\cite{levy2006laplace,vallet2008spectral}.
More importantly,
   the LBO has been successfully applied in high-dimensional data processing.
A well-known application of the LBO in high-dimensional data processing is Laplacian eigenmaps~\cite{belkin2003laplacian}.
To appropriately represent high-dimensional data for machine learning and pattern recognition,
  Belkin et al.~\cite{belkin2003laplacian} proposed a locality-preserving method for non-linear dimensionality reduction using the LBO eigenfunctions.
Laplacian eigenmaps have a natural connection to clustering,
  and can be computed efficiently.

\subsection{Metrics}
In this section,
  we introduce four metrics that are used to quantitatively measure the quality of our index.
\emph{Pearson correlation coefficient} was developed by Karl Pearson~\cite{pearson1895notes}
  from a related idea introduced by Francis Galton in the 1880s~\cite{galton1886regression},
  and its mathematical formula was derived and published by Auguste Bravais in 1844~\cite{bravais1844analyse}.
It is a metric of the linear correlation between two variables $X$ and $Y$,
  which is widely used in financial research,
  medicine,
  bioinformatics,
  climatology,
  signal processing,
  and recommender systems
  ~\cite{de2004measuring,kenett2015partial,mukaka2012guide,zou2016novel,abbot2012application,benesty2008importance,sheugh2015note}.
\emph{Alpha},
  which is also called excess return,
  is a finance term that measures how much a portfolio or investment returns in excess of the market.
It is typically applied in financial research,
  including fund management,
  portfolio analysis,
  leverage aversion,
   and risk parity~\cite{ferson2014alpha,maccheroni2013alpha,asness2012leverage,acharya2016seeking}.
\emph{Beta} originates from the capital asset pricing model(CAPM)~\cite{black1972capital},
  which is a measure of the systematic risk of an individual stock in comparison to the entire market.
Similar to \emph{Alpha},
  \emph{Beta}'s main application area is financial research,
   such as the hedge fund market and stock returns~\cite{dempsey2013capital,zabarankin2014capital,barberis2015x,kisman2015m}.
\emph{Jenson's Alpha} is proposed by Jensen~\cite{jensen1968performance},
  which considers the risk factor \emph{Beta} when calculating excess returns.
It can be seen as a version of the standard \emph{Alpha} based on a theoretical performance index instead of a market index.
\emph{Jensen's Alpha} is also mainly applied to financial research,
  such as portfolio returns,
  market activities analysis,
  stock trading,
   and so on~\cite{kang2013bias,jarrow2013positive,bunnenberg2019jensen,tralvex2013quantative}.

\section{Methodology}

\subsection{Data}\label{sec:pre}
The stock market studied in this paper is the Shanghai Stock Exchange (SSE).
To construct the manifold structure of all stocks in the SSE,
  we use the closing price data from January 2014 to January 2018.
Suppose we want to update the list of constituents at the beginning of the $t$-th year,
  then the price data in the $t-1$-th year is used for research.
If the number of trading days in the $t-1$-th year is $m$,
  the closing price of each stock can be represented by an $m$-dimensional vector $v=(v^1,v^2,\cdots,v^m)$.
Considering listing,
            delisting,
            suspension,
            and other situations,
  the vector of the closing price may be incomplete ($v^i=null$).
Therefore,
   we need to perform some preprocessing,
   to ensure all vectors are complete for the next work.

\textbf{Data completion.}
    When a stock is suspended,
    its closing price may not exist in the database;
    therefore,
    the original closing price vector may be mutilate due to the absence of a few elements.
  In this situation,
    we consider the price of the absent elements to be the same as that of the previous trading day.

\textbf{Data screening.}
  As the study duration is one year,
  there may be some stocks that are listed or delisted at a time node in the target year.
Then,
  the data of these stocks are absent before listing or after delisting.
Completing the absent elements here is inappropriate,
  because the constituents we select are used to represent the entire market in the target year,
  they should be stable in this year,
  at least be traded throughout the year;
  thus,
  such stocks that involve listing and delisting are inappropriate.
Therefore,
  we remove the stocks that are listed or delisted during the study duration.

\textbf{Data transformation.}
In the next section,
  we build connection between stocks using Euclidean distance.
Suppose there are two stocks with the same price change trend and significant price difference,
  then the distance between them will be large.
We aim to avoid this situation,
  because in this paper,
  the connection between stocks is expected to be related only to price changes,
  rather than the prices themselves.
Therefore, this paper adds data transformation in preprocessing.
To decrease the influence of the price,
  we normalized the vectors in the data preprocessing as
   \[
    \tilde{v}^i = \frac{v^i}{\sqrt{\sum_{k=1}^m {(v^k)^2}}}.
  \]

\subsection{High-dimensional data simplification}
After data preprocessing,
   each stock in the studied market is represented by an $m$-dimensional vector,
   which can also be regarded as an $m$-dimensional point.
Meanwhile,
   the entire stock market is regarded as an $m$-dimensional point cloud.
In this section,
  a high-dimensional point cloud simplification method is introduced.
The simplified point cloud is regarded as the constituent stock dataset.

\subsubsection{LBO on manifold}
The Laplace-Beltrami operator $\Delta$ is the divergence of the gradient.
Suppose $(M,g)$ is a Riemannian manifold
   (a differentiable manifold $M$ with Riemannian metric $g$),
   and $f\in C^{2}$ is a real-valued function defined on $M$.
The LBO $\Delta$ on $M$ is defined as
  \begin{equation}\label{eq:lbo_org}
    \Delta f = div(grad\ f),
  \end{equation}
  where $grad\ f$ is the gradient operator of $f$,
    and $div(\cdot)$ is a divergence operator.

The eigenvalues and eigenfunctions of the
    Eq.~\ref{eq:lbo_org} can be calculated by solving the eigen-equation:
    \begin{equation}\label{eq:helmholtz}
    \Delta f = -\lambda f,
    \end{equation}
  where $\lambda$ is a real number.
  The solution of Eq.~\ref{eq:helmholtz} is a list of eigenvalues such as
  \begin{equation}\label{eq:eigen_value}
    0\leq\lambda_{0}\leq\lambda_{1}\leq\lambda_{2}\leq...\leq+\infty{},
  \end{equation}
  with the corresponding eigenfunctions
  \begin{equation}\label{eq:eigen_function}
        \phi_0(x), \phi_1(x), \phi_2(x), \cdots,
  \end{equation}
    which are normalized and orthogonal.
In the case of a closed manifold,
    the first eigenvalue $\lambda_{0}$ is always zero.

\begin{figure}[!htbp]
   \centering
   \subfigure[]
   {
        \includegraphics[width=0.7\textwidth]
            {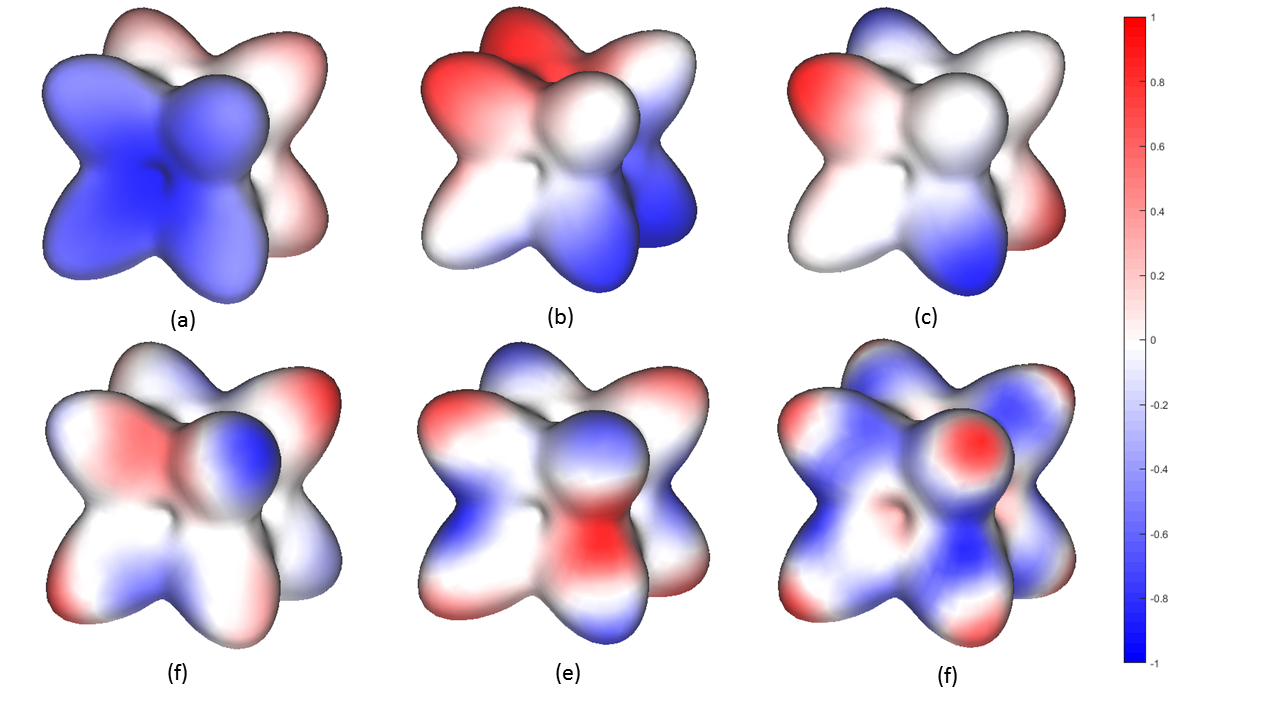}
   }
   \caption
   {
        Overview of LBO eigenfunctions with different frequencies of a 2D manifold.
   }
   \label{fig:spectrum}
 \end{figure}

The eigenvalues $\lambda_i, i=0,1,2,\cdots$
    specify the discrete frequency domain of an LBO,
    and the eigenfunctions are the extensions of the basis functions in the Fourier analysis to a manifold~\cite{vallet2008spectral}.
The numerical distributions of different eigenfunctions correspond
    to different frequency geometric information,
    and the eigenfunctions with larger eigenvalues contain higher frequency information~\cite{Shen2006Spectral}.
In Fig.~\ref{fig:spectrum},
  the eigenfunctions of a 2D manifold are demonstrated,
  their eigenvalues are increasing from Fig.~\ref{fig:spectrum}(a) to Fig.~\ref{fig:spectrum}(f).
For each eigenfunction,
  the value is shown in colors from blue to red,
  corresponding to -1 to 1,
  respectively.
The dark red or blue parts correspond to the geometric features of the model,
  such as convex areas and concave areas.
This provides the main inspiration for our constituent selection method.
We detect the local extreme points on eigenfunctions
  (typically the points with the darkest color),
  and merge them into the simplified data set,
  which can be regarded as the constituents of the MF index.
In the next section,
  the specific details are introduced.


\subsubsection{Discrete LBO on high-dimensional dataset}\label{sec:disLBO}
Our research object is all the stocks in the Shanghai Stock Exchange in a target year.
We use a daily closing price curve to describe a stock.
After the data preprocessing mentioned in Section~\ref{sec:pre},
  each stock can be represented as a high-dimensional vector.
According to the assumption in ~\cite{balasubramanian2002isomap,donoho2003hessian,belkin2003laplacian},
  the stocks are points distributed in a lower-dimensional manifold that is embedded in a higher-dimensional Euclidean space.
This section describes the construction of the manifold structure and discrete LBO
  using these discrete high-dimensional points.

Suppose the number of stocks in the SSE in the study year is $n$,
  each stock is represented as a high-dimensional vector $v_i=(v_i^1,v_i^2,\cdots,v_i^m), i=1,2,\cdots,n$.
The dimension of each stock $v_i$ is $m$,
  the number of trading days in this year,
  so the vector $v_i$ can also be regarded as a point in the $m$-d Euclidean space.
We use the discretization method developed in~\cite{belkin2003laplacian} to construct the discrete LBO.
The LBO discretization method consists of two steps:
   \begin{enumerate}
      \item  adjacency graph construction,
      \item  weight matrix construction,
    \end{enumerate}
   which are explained in detail as follows.

\textbf{Adjacency graph construction.}
  In our implementation,
    we employ $k$-nearest neighbors (\emph{KNN}) to construct an adjacency graph (a manifold connection structure):
  For any vertex $v$,
    its \emph{KNN} set $N_{v}=\{v_i, i=1,2,\cdots,k\}$ is determined using KD-tree~\cite{muja2009flann},
    which is a fast algorithm to find $k$-nearest points of $v$ in high-dimensional Euclidean space.
  Then the connection between the vertex $v$ and each neighbor
    $v_i, i=1,2,\cdots,k$ is established,
    and thus constructing the adjacency graph.
  One thing worth noting is that the \emph{KNN} are not symmetric for sure,
    i.e., it is possible that the vertex $v_i \in N_{v_j}$, 
    yet, $v_j \notin N_{v_i}$,
    then the weight matrix (see below) constructed by the \emph{KNN} method is not guaranteed to be symmetric.

\textbf{Weight computation.}
  Using the adjacency graph for each point $v_i$ in the given data set,
    the connections between point $v_i$ and the other points in the data set are established.
  Based on the adjacency graph,
    the weight $w_{ij}$ between point $v_i$ and $v_j$ can be calculated as follows:
     \begin{equation}\label{eq:weight_element}
      w_{ij}=
     \left\{
             \begin{array}{lr}
             -e^{-\frac{\norm{v_{i}-v_{j}}_2^2}{t}},      &$if $i,j$ are adjacent,$ \\
              \sum_{k \neq i} -w_{ik} ,     &$if $ i = j, \\
              0,   &$otherwise$,\\
             \end{array}
    \right.
    \end{equation}
    where $t$ is an adjustment parameter.

  As stated above, the weight matrix $\tilde{W}=[w_{ij}]$
    \pref{eq:weight_element} constructed by the \emph{KNN} method is not symmetric,
    so we take the following matrix:
    \[
        W = \frac{\tilde{W} + \tilde{W}^T}{2}
    \]
    as the weight matrix,
    which is a symmetric matrix.
  Moreover, we construct a diagonal matrix
  \[
        A = diag(a_1, a_2, \cdots, a_n),
  \]
    where $a_i = w_{ii}$ (refer to Eq.~\pref{eq:weight_element}).

Finally,
  the LBO on manifold can be discretized into the matrix
  \[
        L = A^{-1} W,
  \]
  whatever the dimension of the space the LBO defined on.
   Meanwhile, the Eq.~\ref{eq:helmholtz} is discretized as
  \[
        L \phi = A^{-1} W \phi = \lambda \phi,
  \]
    where $\lambda$ is the eigenvalue of the Laplace matrix $L$,
    and $\phi$ is the corresponding eigenvector.
  It is equivalent to
  \begin{equation}\label{eq:discrete_helmholtz}
    W \phi =\lambda A \phi.
  \end{equation}
  Because the matrix $W$ is symmetric,
    and the matrix $A$ is diagonal and positive,
    it has been shown~\cite{belkin2003laplacian} that the eigenvalues $\lambda$ are all non-negative real numbers satisfying
    \[
        0=\lambda_1 \leq \lambda_2 \leq \cdots \leq \lambda_n,
    \]
  and the corresponding eigenvectors $\phi_i, i=1,2,\cdots,n$
    are orthogonal,
  i.e.,
  \[
    <\phi_i, \phi_j>_A = \phi_i^T A \phi_j = 0,\ i \neq j.
  \]

\subsubsection{Constituent selection}\label{sec:simplification}
 By solving Eq.~\pref{eq:discrete_helmholtz},
    eigenvalues $\lambda_1 \leq \lambda_2 \leq \cdots \leq \lambda_n$,
    and the corresponding eigenvectors $\phi_i, i=1,2,\cdots,n$ are determined.
 The local maximum and minimum of the eigenvectors
    are used as the feature points.
 As stated above,
    the eigenvalue $\lambda_i$ measures the frequency of its corresponding eigenvector $\phi_i$.
The larger the eigenvalue,
    the higher is the frequency of the eigenvector $\phi_i$,
    and the greater is number of feature points in the eigenvector $\phi_i$.
 The data simplification algorithm first detects the feature points of the
    eigenvector $\phi_1$,
    and adds them to the simplified dataset.
 Then,
    it detects and adds the feature points of $\phi_2$ to the simplified dataset.
 This procedure is performed iteratively
    until the number of elements of the simplified data reaches the required number of constituents.

As eigenvector $\phi_i, i=1,2,\cdots,n$ is a scalar
    function defined on an unorganized point set in a high-dimensional space,
    the detection of the maximum and minimum of the defined function is not straightforward.
 For each point $x$,
    we use its \emph{KNN} results $N_x$,
    calculated previously in Section~\ref{sec:disLBO}
    as the neighbor construction.
 The method for detecting the maximum and minimum of the function $\phi(x)$ is elucidated as follows:

 \textbf{Maximum and minimum point detection:}
  For a data point $x$ in the data set,
    if $\phi(y) < \phi(x)$, $\forall y \in N_x$,
    the data point $x$ is a local maximum point of $\phi(x)$.
  Conversely, if $\phi(y) > \phi(x)$,  $\forall y \in N_x$,
    the data point $x$ is a local minimum point of $\phi(x)$.

\begin{figure}[!htbp]
   \centering
   \subfigure[]
   {
        \includegraphics[width=0.3\textwidth]
            {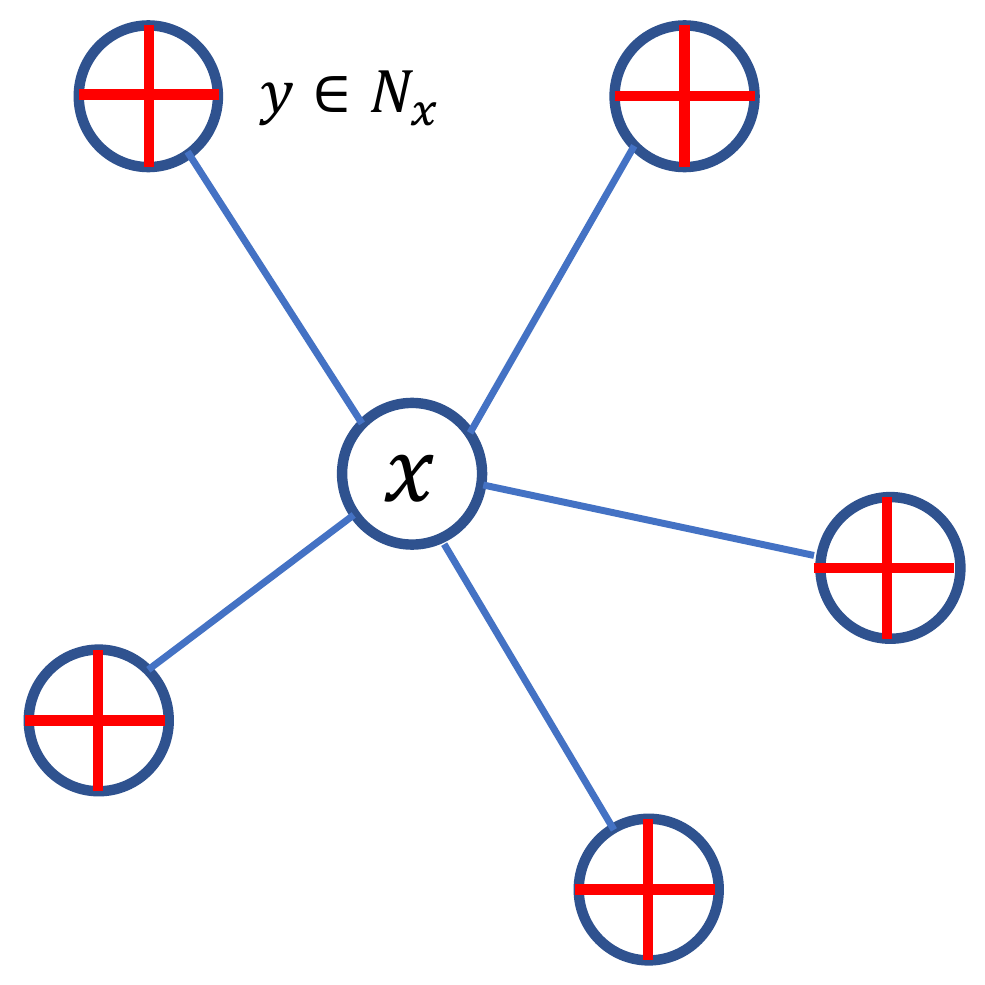}
   }
   \subfigure[]
   {
        \includegraphics[width=0.3\textwidth]
            {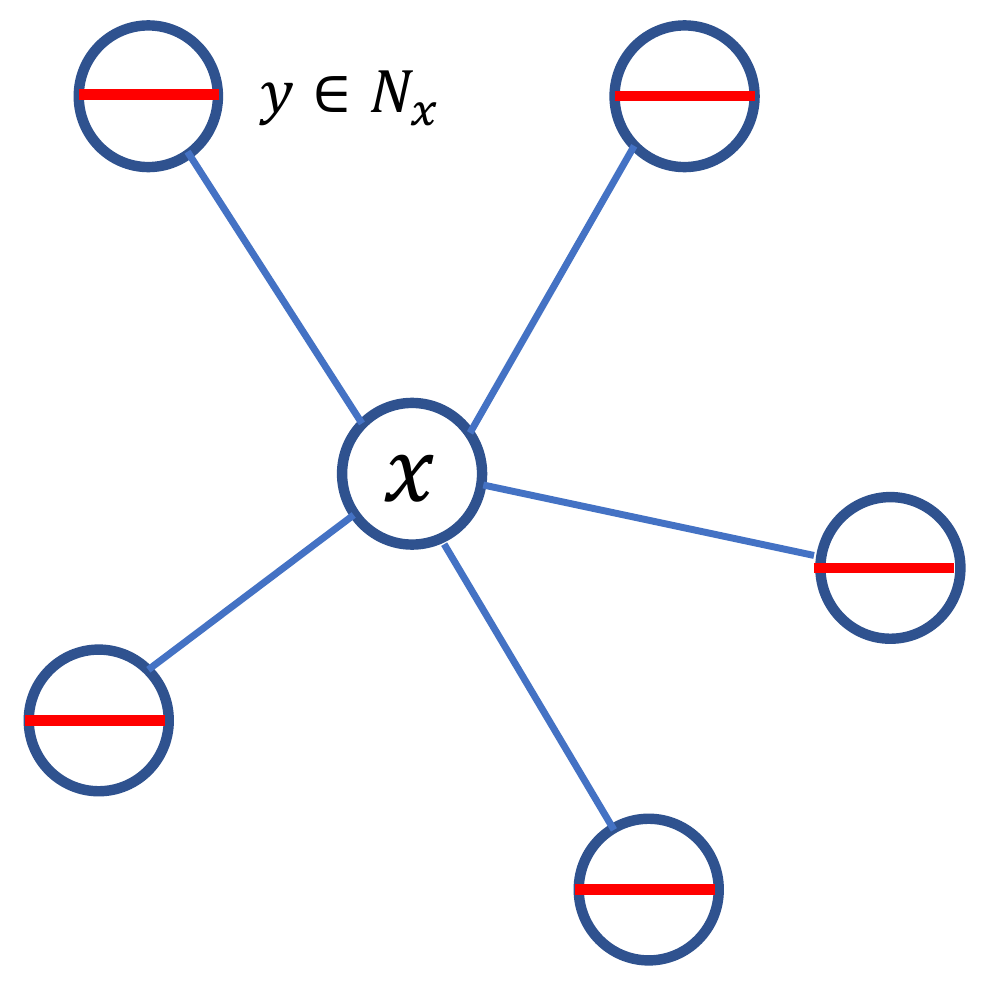}
   }
   \subfigure[]
   {
        \includegraphics[width=0.3\textwidth]
            {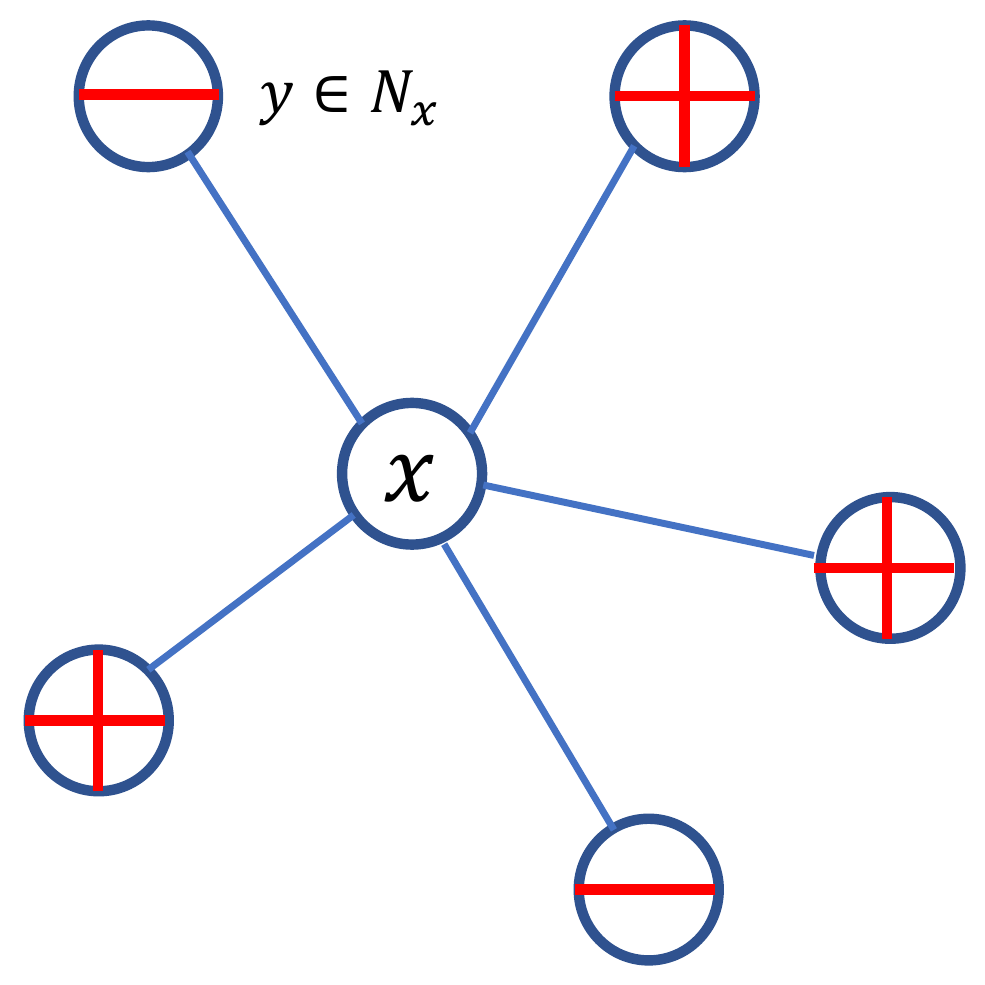}
   }
   \caption
   {
        Maximum and minimum point detection:
        $x$ is a local maximum point in (a), a local minimum point in (b), and neither in (c).
   }
   \label{fig:maxmin}
 \end{figure}
As illustrated in Fig.~\ref{fig:maxmin},
  for any $y \in N_x$,
  if $\phi(y) < \phi(x)$,
    it is labeled as $\oplus$;
    otherwise, it is labeled as $\ominus$.
For a local maximum point,
  all the neighbors are labeled as $\oplus$;
for a local minimum point,
  all the neighbors are labeled as $\ominus$;

As each data point $x$ corresponds to a stock in the study stock market,
  the simplified dataset composed of feature points naturally corresponds to a simplified stock set,
  which is regarded as the set of constituents to compute the MF index.

\subsection{Index computation}
In this section,
  a detailed methodology to generate the MF index is introduced.

For stock indexes related to the SSE,
    the SSE composite index is composed of all SSE listed stocks,
    including A share and B share stocks.
For other indexes such as SSE 50, SSE 100, and SSE 180,
    their constituents are selected from a larger constituent space.
Taking the SSE 50 as an example,
       its constituents are selected from the constituent list of SSE 180.
The selection criteria include size, liquidity and so on.
After ranking the stocks by negotiable market capitalization and trading value,
   in principle,
   the top 50 ranked stocks are selected except for stocks with abnormal market performance,
     and some inappropriate stocks will be removed by the Index Advisory Committee.
In our method,
   the constituents of the MF index are determined by the high-dimensional data simplification method stated in Section~\ref{sec:simplification},
   and the selected constituent stock set is regarded as a simplified dataset which can represent the entire SSE stock market.

The index calculation method we use is the same as that of the SSE composite index.
   The index is weighted using the following formula:
    \begin{equation}\label{eq:SSE CI}
      {MF(t)} = \frac{\sum{P_i(t) N_i}}{D} {B},
    \end{equation}
     where $MF(t)$ is the index value at time $t$.
   For the $i$-th stock in the database,
     $P_i(t)$ is the stock price at time $t$,
     $N_i$ is the number of shares issued.
   $B$ is the base level, which is a fixed number.
   $D$ is a divisor which is used to adjust the MF index
     when changes occur in the constituent list (delisting),
      or the share structure (share changes),
     or market cap (right issue and bonus issue) due to non-trading factors.
  When such situations happen to a stock,
     its market cap ($P_i(t) \times N_i$) may be changed;
     then, the value of MF index changes,
     but we expect the index to be unchanged because it should be affected only by trading.
   This is the reason for setting divisor D in the calculation formula,
     and the detailed adjustment formula of the new divisor $D_{new}$ is
     \begin{equation}\label{eq:divisor}
       D_{new}=D_{old} \frac{M_{new}}{M_{old}},
     \end{equation}
    where $D_{old}$ is the divisor before adjustment,
    $M_{old}$ is the market cap before adjustment
    and $M_{new}$ is the market cap after adjustment.

At the end of this section,
     the generation method of the MF index is presented in Algorithm~\ref{alg:MFIndex}.
  \begin{algorithm}
  \caption{MF Index}
  \label{alg:MFIndex}
  \begin{algorithmic}
  \STATE {\textbf{Input}: Closing price and market cap of stocks in SSE,
                          number of constituents $N$};
  \STATE{1. Data preprocessing: the stocks are represented as $v_i, i=1,2,\cdots,n$};
  \STATE{2. Build the \emph{KNN} of all stocks, construct the discrete LBO $L$};
  \STATE{3. Solve Eq.~\ref{eq:discrete_helmholtz}, obtain eigenvectors $\phi_i$};
  \STATE{4. Simplified dataset $S={\O}$, k=0};
  \STATE{5. Data simplification}:
  \WHILE{$|S|<N$}
        \STATE{$k=k+1$};
        \STATE{Detect the feature points on $\phi_k$, add them into $S$};
 \ENDWHILE
 \STATE{if $|S|>N$ delete the stock which have the smallest market cap until $S=N$};
  \STATE{6. Calculate the MF index using the constituents in $S$ using Eq.~\ref{eq:SSE CI}}.
 \STATE {\textbf{Output}: MF N Index.}
   \end{algorithmic}
   \end{algorithm}

\section{Metrics}\label{sec:metric}
As stated above,
  after selecting the constituents using our high-dimensional data simplification method,
  and computing using the weighted formula Eq.~\ref{eq:SSE CI},
  the proposed MF index is generated.
The issue now is how the quality of the MF index should be evaluated.
In this section,
  four metrics are devised to measure the MF index.
They are \emph{Pearson correlation coefficient},
  \emph{Alpha},
  \emph{Beta}
  and \emph{Jensen's alpha}.
They measure the quality of the MF index from two aspects:
 \begin{itemize}
    \item data correlation(\emph{Pearson correlation coefficient}),
    \item risk premium(\emph{Alpha}, \emph{Beta} and \emph{Jensen's alpha}).
  \end{itemize}


\emph{Pearson correlation coefficient(Pearson):}
The direct purpose of designing a stock index is
  to reflect the overall price activity of the stock market by using fewer stocks.
The constituents we select are expected to represent the entire stock market
  better than other indexes with the same constituent number.
For example,
  if we set the number of constituents of the MF index to 50,
  then the MF 50 should be more similar to the SSE composite index than the SSE 50.

We use \emph{Pearson} to describe the similarity between two indexes.
In statistics,
  the \emph{Pearson} is a metric of the linear correlation between two variables $X$ and $Y$.
In this paper,
  two indexes are represented as high-dimensional vectors $v_X$ and $v_Y$,
  dimension $m$ is the number of trading days in a target study year,
the vector $v_X=(v_X^1,v_X^2,\cdots,v_X^m)$ can be viewed as $m$ samples of variable $X$,
  and vector $v_Y=(v_Y^1,v_Y^2,\cdots,v_Y^m)$ is the $m$ sample of variable $Y$.
Then,
   the \emph{Pearson} can be calculated as follows:
   \begin{equation}\label{eq:pcc org}
     \rho_{X,Y}=\frac{cov(X,Y)}{\sigma_X\sigma_Y}.
   \end{equation}
If we denote
   \begin{equation*}
      \bar{v}_X = \frac{\sum_{k=1}^m v_X^k}{m}
   \end{equation*}
   and
   \begin{equation*}
      \bar{v}_Y = \frac{\sum_{k=1}^m v_Y^k}{m},
   \end{equation*}
   Eq.~\pref{eq:pcc org} can be calculated by
   \begin{equation}\label{eq:pcc cal}
     \rho_{X,Y}= \frac{\sum_{k=1}^m (v_X^k-\bar{v}_X)(v_{Y}^k-\bar{v}_Y)}
     {\sqrt{\sum_{k=1}^m (v_X^k-\bar{v}_X)^2}
     \sqrt{\sum_{k=1}^m (v_Y^k-\bar{v}_Y)^2}}.
   \end{equation}
\emph{Pearson} measures the correlation between vectors $v_X$ and $v_Y$.
The closer the \emph{Pearson} is to 1,
  the higher is the correlation between the two indexes $v_X$ and $v_Y$.

However,
  in terms of investment fund management,
  because an index can be regarded as a portfolio,
  and a well-performing index should be instructive for investors to make asset portfolios compared to other indexes,
  we measure the MF index from the perspective of the risk premium.
In this section,
  we use \emph{Alpha},
         \emph{Beta} and \emph{Jensen's Alpha},
  which are three metrics to measure the quality of the MF index,
  from three investment perspectives,
       the excess returns,
       the risk,
       and the excess returns after the risk adjustment,
       respectively.

Before introducing the metrics,
  we provide the following symbol descriptions:
   \begin{itemize}
    \item $R_i$: the realized return of the portfolio or investment,
    \item $R_m$: the realized return of the appropriate market index,
    \item $R_{rf}$: the risk-free rate of return.
   \end{itemize}
In this paper,
  all returns are monthly returns,
  $R_i$ is calculated using our MF index,
  $R_m$ is calculated using the SSE composite index (SSE) during the same period,
  and $R_{rf}$ is approximated using national debt and $R_{rf}=0.2\%$ for easy calculation.


\emph{Alpha}($\alpha$) is a financial term that measures how much a portfolio or investment returns in excess of the market.
In investment fund management,
  \emph{Alpha} is commonly calculated as the excess returns a fund manager achieves over a fund's stated benchmark.
In this paper,
  we calculate the average of the MF index's excess monthly returns over a specific benchmark,
  with the average of the SSE's monthly returns,
  as:
   \begin{equation}\label{eq:alpha}
     \alpha = \bar{R}_i - \bar{R}_m.
   \end{equation}
If the MF index has an $\alpha>0$,
  the realized monthly returns of the MF index exceeds the SSE,
  which means the MF index is a portfolio that performs better than the SSE in the excess monthly returns.
Meanwhile,
  \emph{Alpha} can also measure the similarity between the index and the market,
  that is,
  the closer \emph{Alpha} is to 0,
  the higher is the similarity between the index and the market.
\emph{Beta}($\beta$) is a measure of the systematic risk of an individual stock in comparison to the entire market.
In statistical terms,
  it represents the slope of the line through a regression of data points from an index's returns against the entire market.
In this paper,
   we calculate $\beta$ as:
   \begin{equation}\label{eq:beta}
     \beta =\frac{cov(R_i,R_m)}{var(R_m)}.
   \end{equation}
If the MF index has a $\beta=1.0$,
  it indicates that its price activity is strongly correlated with the market (SSE).
If the MF index has a $\beta<1.0$,
  it means that the security of the MF index is theoretically less volatile than the market.
If the MF index has a $\beta>1.0$,
  it indicates that the security of the MF index is theoretically more volatile than the market.
Some stocks even have negative $\beta$,
  a $\beta = -1.0$,
   which means that the index is inversely correlated to the market benchmark,
  as if it were an opposite mirror image of the benchmark’s trends.
Meanwhile,
  \emph{Beta} can also measure the similarity between the index and the market,
  that is,
  the closer \emph{Beta} is to 1.0,
  the higher is the similarity between the index and the market.

\emph{Jensen's Alpha}($\alpha_J$) is a version of the standard \emph{Alpha} based on a theoretical performance index instead of a market index.
The theoretical performance index usually needs to be adjusted by a risk factor.
In this paper,
  we use $\beta$ to calculate $\alpha_J$ as:
   \begin{equation}\label{eq:Jalpha}
     \alpha_J = R_i-[R_{rf}+\beta(R_M-R_{rf})].
   \end{equation}
If the realized returns of the MF index exceeds over the theoretical performance index after being adjusted by beta,
  i.e. $\alpha_J>0$,
  the index can be said to have excess returns.
Investors are always looking for investment products with higher $\alpha_J$ values.
Similar to $\alpha$,
  $\alpha_J$ can also measure the similarity between the index and the market,
  the closer $\alpha_J$ is to 0,
  the higher is the similarity between the index and the market.

\section{Implementation, results and discussions}
The market under study is the SSE.
A total of 1559 stocks were extracted within a period of 5 years,
  from 2014 to 2018.
We update the constituent list of the MF index on the first trading day of each year,
   that is,
   we use the closing price of all stocks in the $t$-th year to determine the constituents,
  and compute the MF index in the $t+1$-th year using these constituents.
The excess returns mentioned above are monthly returns.
We use the SSE composite index as the market benchmark,
  and national debt to approximate risk-free returns.

The proposed method is implemented using C++ with OpenCV and ARPACK~\cite{lehoucq2007arpack},
    and it executes on a PC with a 3.6 GHz Intel Core i7 4790 CPU,
    GTX 1060 GPU,
    and 16GB memory.
We employed the OpenCV function 'cv::flann' to perform the \emph{KNN} operation~\cite{muja2009flann}
    and ARPACK to solve the sparse symmetric matrix eigen-equation(Eq.~\ref{eq:discrete_helmholtz}).

\subsection{Data approximation}
 \begin{figure}[!htbp]
   \centering
   \subfigure[]
   {
        \includegraphics[width=1.0\textwidth]
            {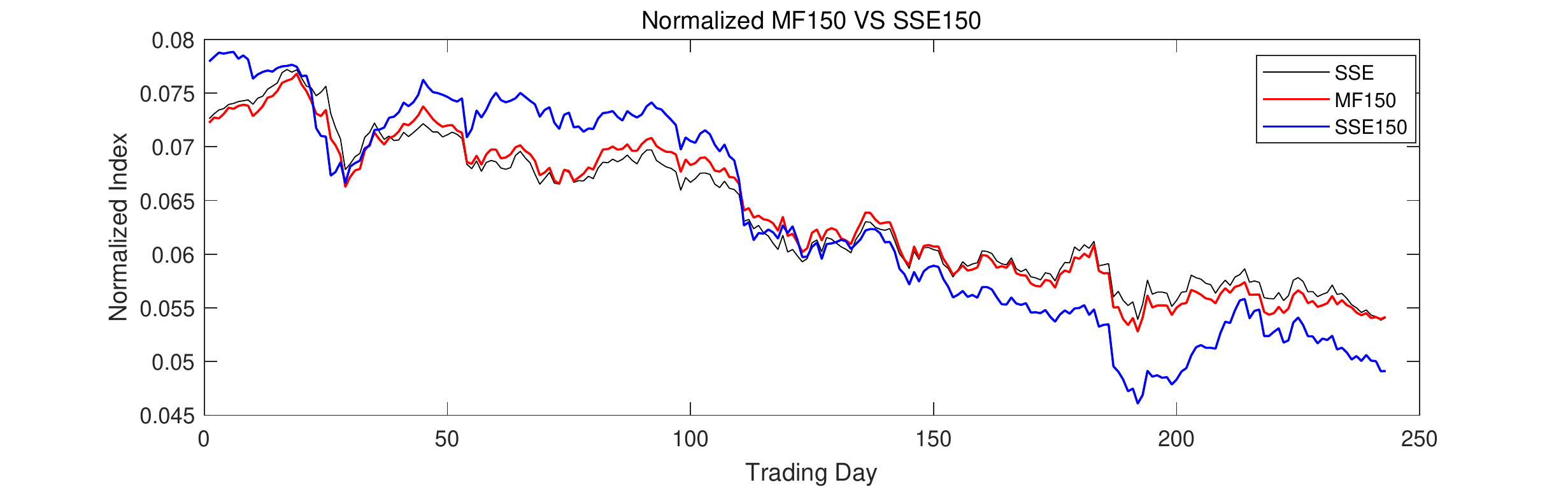}
   }
   \subfigure[]
   {
        \includegraphics[width=1.0\textwidth]
            {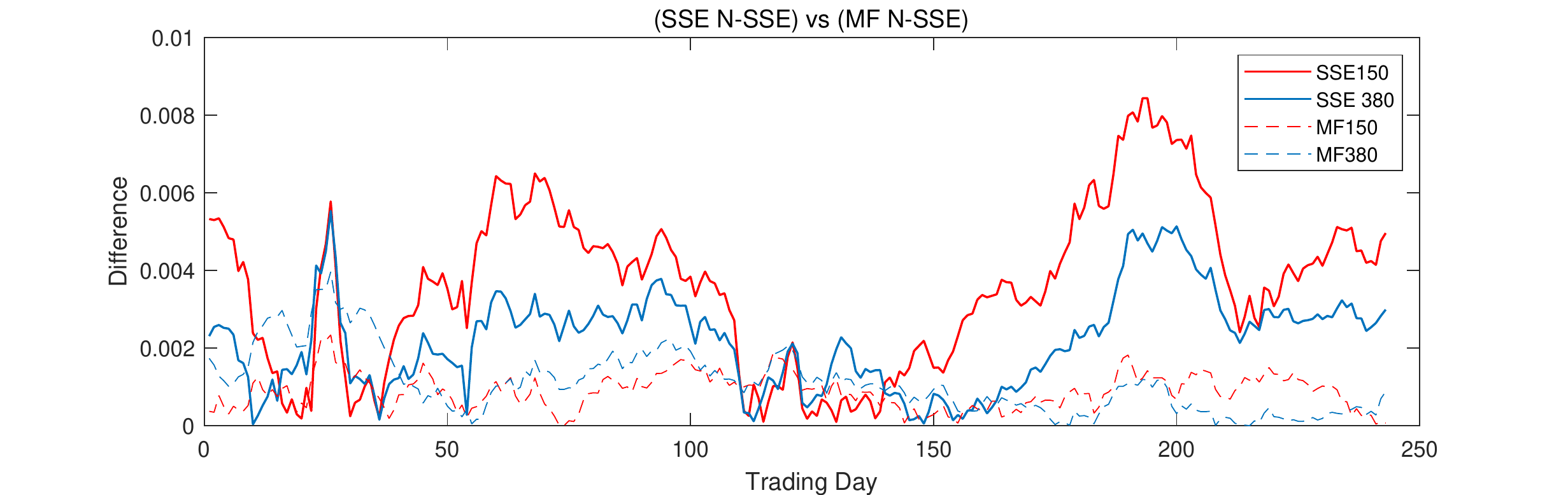}
   }
   \subfigure[]
   {
        \includegraphics[width=1.0\textwidth]
            {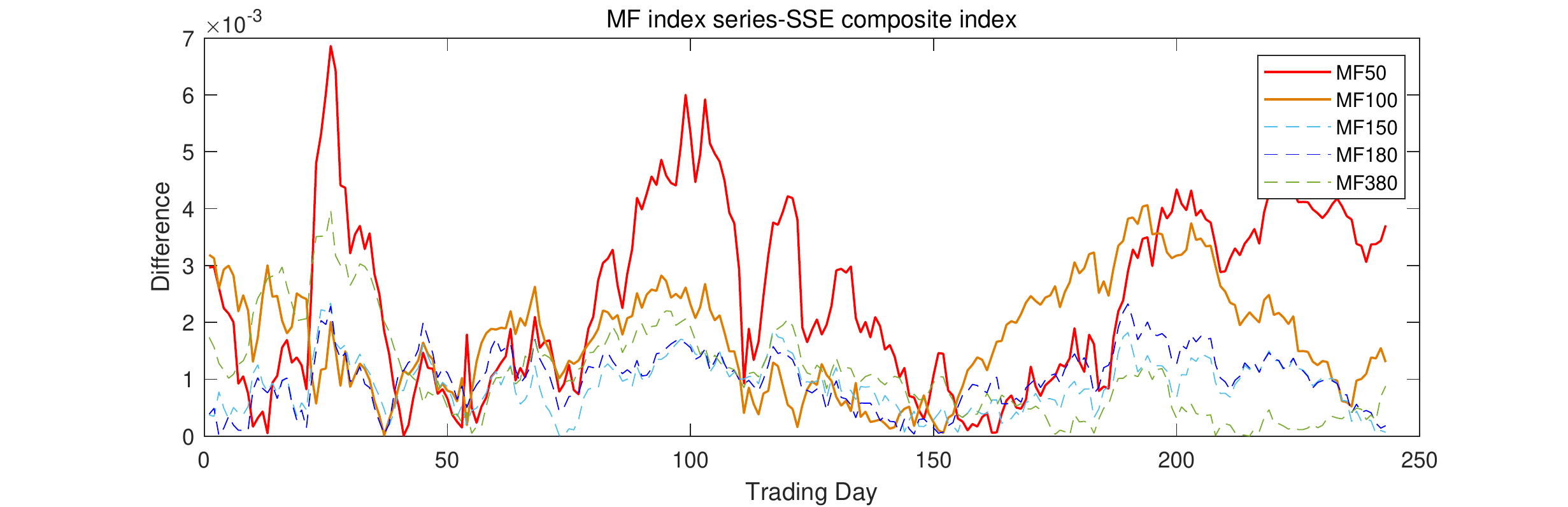}
   }
     \subfigure[]
   {
        \includegraphics[width=1.0\textwidth]
            {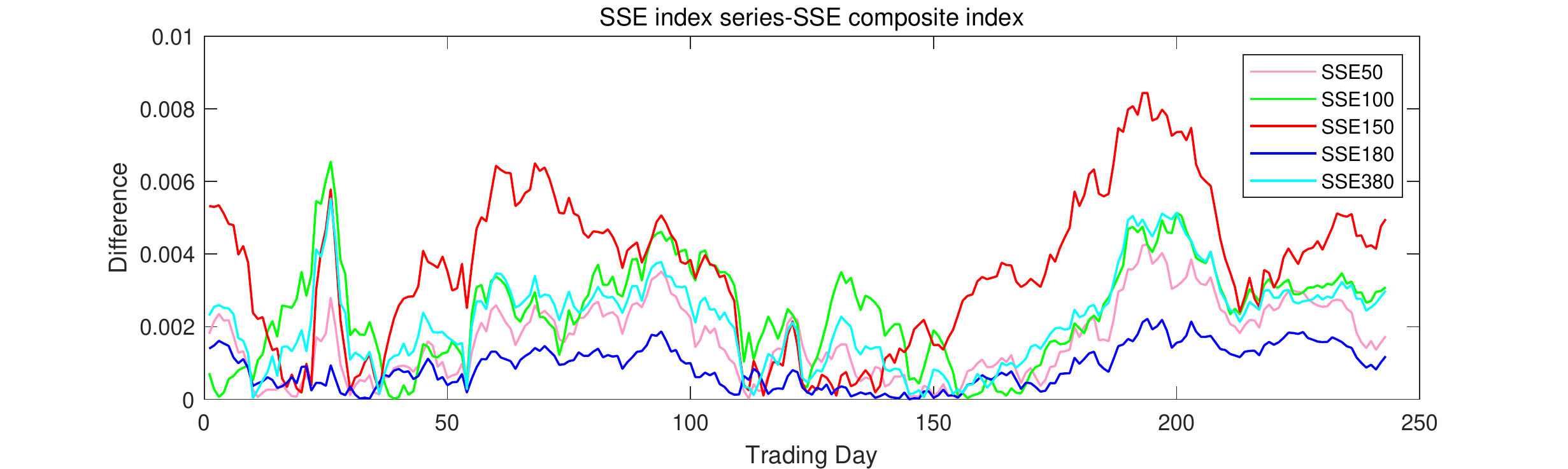}
   }
   \caption
   {
        Display of MF index series,
          SSE index series,
          and SSE composite index in 2018.
        (a)Normalized MF 150, SSE 150, and SSE.
        (b)The difference between MF 150 (180) and SSE vs.
           that between SSE 150 (180) and SSE.
        (c)The difference between the MF index series and SSE.
        (d)The difference between the SSE index series and SSE.
   }
   \label{fig:SSEVSMF2018}
 \end{figure}
Firstly,
  in Fig.~\ref{fig:SSEVSMF2018},
  a visual display of the MF index series is demonstrated to illustrate the approximation of the MF index to the market (SSE).
We extract the closing prices of all stocks in 2017 (244 trading days),
  and perform preprocessing mentioned in Section~\ref{sec:pre},
  and then,
  a 244-d point cloud set including 1438 points is built.
The MF index generation algorithm~\ref{alg:MFIndex} sets the number of constituents as
  50, 100, 150, 180 and 380,
  for later comparison with SSE 50, SSE 100, SSE 150, SSE 180 and SSE 380, respectively.
Using the closing price and market cap data in 2018,
  the proposed MF indexes,
      MF 50, MF 100, MF 150, MF 180, and MF 380 are generated.

In Fig.~\ref{fig:SSEVSMF2018}(a),
  the normalized MF 150 and its comparisons,
  SSE and SSE 150 are demonstrated.
Remarkably result in Fig.~\ref{fig:SSEVSMF2018}(a) shows that
   the red line is closer to the black line than the blue line,
   which means that MF 150 (red line) is more approximate to the SSE composite index (black line)
   than the SSE 150 (the blue line).
To show the meaning of approximate more intuitively,
   we calculate the absolute value of the difference between the constituent indexes and the SSE composite index.
In Fig.~\ref{fig:SSEVSMF2018}(b),
  the differences between the MF index series (MF 150, MF 380) and the SSE
  are demonstrated,
  which are drawn as dotted lines.
As comparisons,
  the differences between the SSE index series (SEE 150, SSE 380) and the SSE are drawn as solid lines.
The results show that the values of the solid lines are almost larger
  than those of the dotted lines in the same color,
  which means that the distance between SSE 150 (380) and SSE is larger than the distance between MF 150 (380) and SSE.
This is summarized as the MF index series that can more precisely reflect market price activity than the SSE index series.

In addition,
  the comparison of the MF indexes with different number of constituents is demonstrated in Fig.~\ref{fig:SSEVSMF2018}(c).
Lines with fewer constituent numbers (MF 50 and MF 100, colored in red and orange)
   have larger difference values than others.
This result means
  that the MF index approximates to the SSE composite index as the number of constituents increases.
This corresponds to the theory in high-dimensional data simplification in Section~\ref{sec:simplification}.
With increasing constituent numbers,
  our method detects feature points on eigenvectors corresponding to higher eigenvalues,
  which contains higher frequency information of the manifold,
  and the simplified dataset is gradually approaching to the entire dataset.
This result does not hold in that of the SSE index series,
  in Fig.~\ref{fig:SSEVSMF2018}(d),
  as the lines do not show any relationship between the values and the number of constituents.

 \begin{figure}[!htbp]
   \centering
   \subfigure[]
   {
        \includegraphics[width=0.4\textwidth]
            {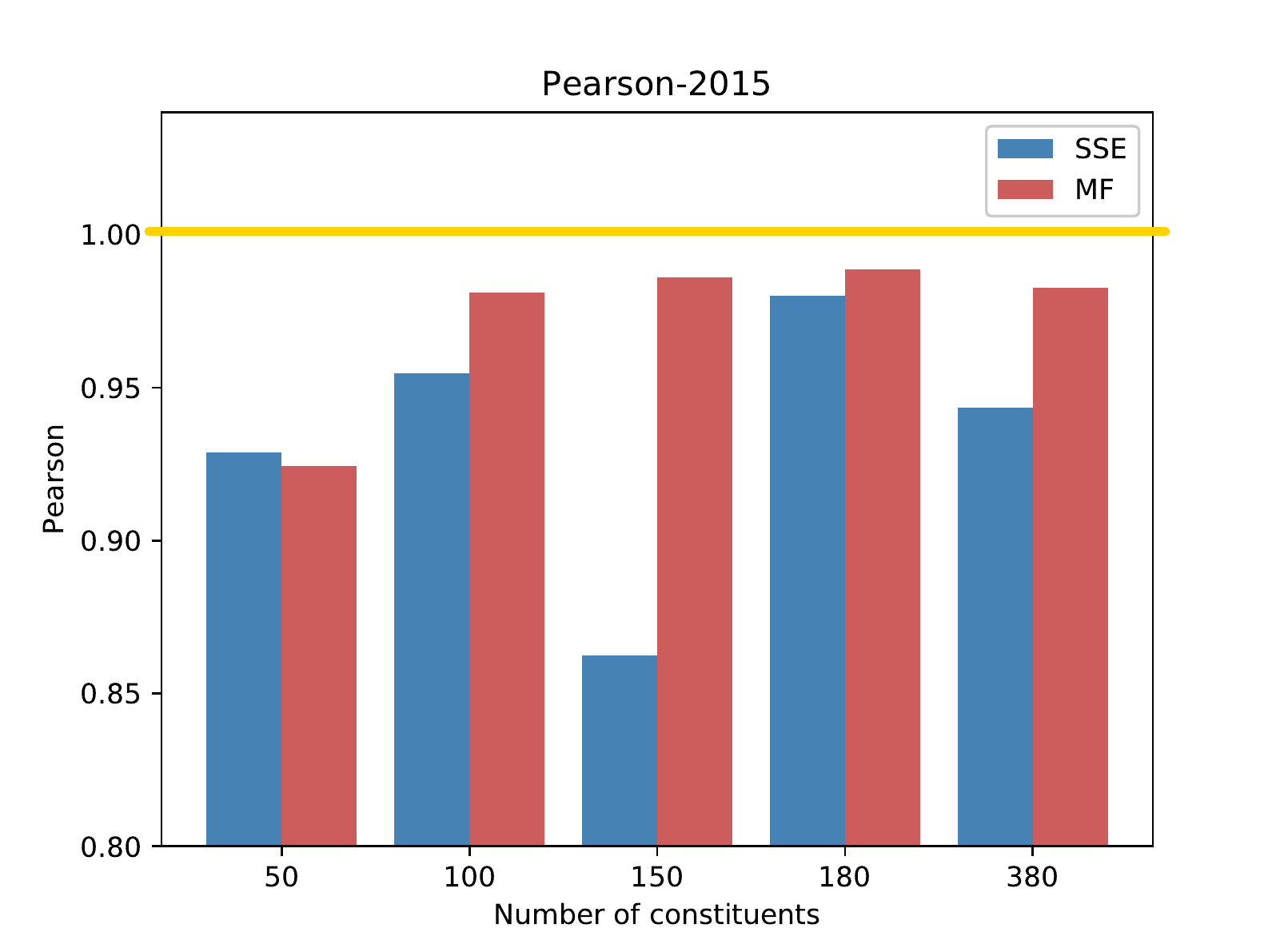}
   }
   \subfigure[]
   {
        \includegraphics[width=0.4\textwidth]
            {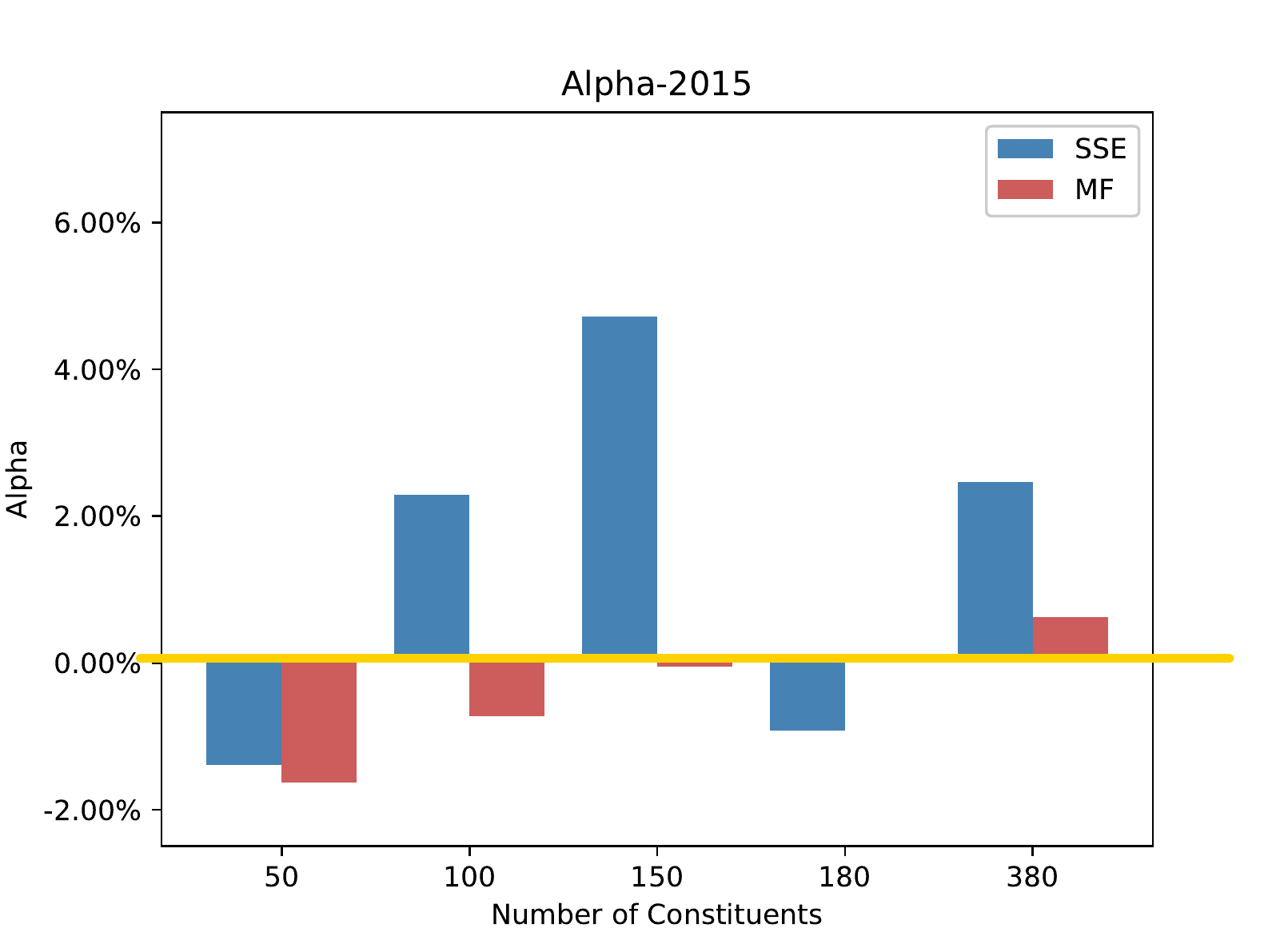}
   }
      \subfigure[]
   {
        \includegraphics[width=0.4\textwidth]
            {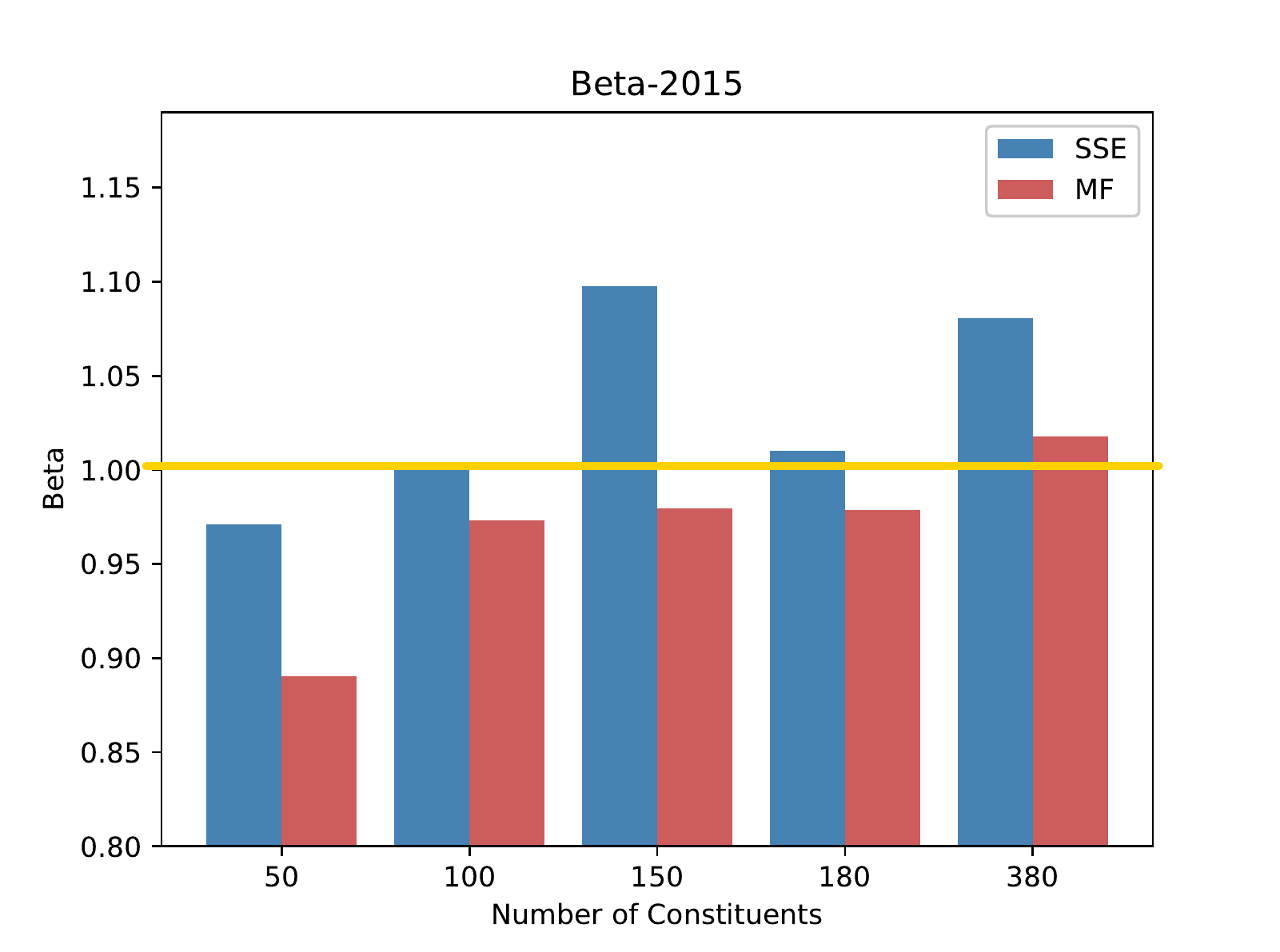}
   }
    \subfigure[]
   {
        \includegraphics[width=0.4\textwidth]
            {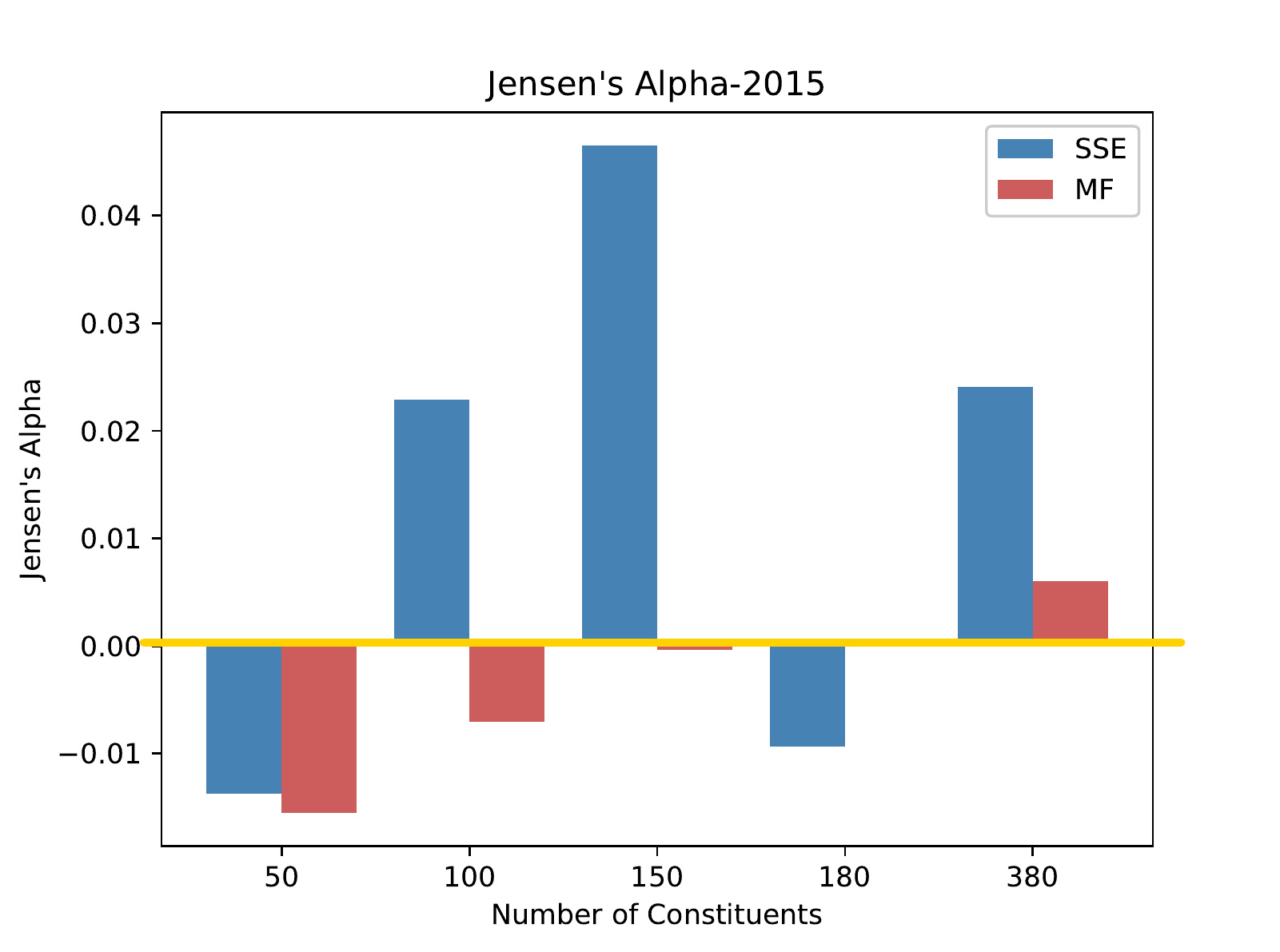}
   }
   \caption
   {
        Pearson, Alpha, Beta and Jensen's Alpha of indexes in 2015.
   }
   \label{fig:4in2015}
 \end{figure}

\subsubsection{Metrics}\label{metric:reaults}
This section quantitatively demonstrates the characteristics of the MF index series
  using the four metrics mentioned in Section~\ref{sec:metric}.
In Fig.~\ref{fig:4in2015},
  we calculate $Pearson$,
  $Alpha$,
  $Beta$ and $Jensen's Alpha$ of the MF index series and the SSE index series in 2015.

\emph{Pearson correlation coefficient} is a quantitative performance of the approximation
  and reflects the similarity between the indexes and the stock market.
We set a base line(yellow) at 1.0;
  the closer the \emph{Pearson} is to 1.0,
  the higher is the correlation between the indexes and the market (SSE).
In Fig.~\ref{fig:4in2015}(a),
  the $Pearson$ of MF 100, MF 150 MF 180, and MF 380
  are closer to the base line than those of SSE 100, SSE 150 SSE 180, and SSE 380.
This result indicates that MF 100, MF 150, MF 180, and MF 380
  perform more approximately to the stock market (SSE)
  than SSE 100, SSE 150, SSE 180, and SSE 380,
  respectively.

\emph{Alpha} calculates the excess monthly returns of indexes over a benchmark(SSE composite index).
If the \emph{Alpha} of the index is 0,
  the index have the same excess returns as the market(SSE).
We set the base line at 0;
  the closer \emph{Alpha} is to 0,
  the higher is the similarity between the index and the market.
In Fig.~\ref{fig:4in2015}(b),
  the \emph{Alpha} of MF 100, MF 150, MF 180, and MF 380
  are closer to 0
  than those of SSE 100, SSE150, SSE 180, and SSE 380,
  respectively.
This result means,
   the returns of the MF index series are closer to the market than those of the SSE indexes.
From the perspective of \emph{Alpha},
  the MF index series is more approximate to the stock market than the SSE index series (The Alpha of MF 180 and the Jensen's Alpha of MF 180 are close to 0,
  so they are difficult to recognize).
To reinforce this conclusion,
   we calculate the $Alpha$ of each index in four years,
   and display the results in Table~\ref{tbl:alpha}.
In order to make it easier to judge which is closer to the market,
   we marked the $Alpha$ that are closer to 0.
In Table~\ref{tbl:alpha},
   the $\alpha$ of MF indexes are close to 0 than those of the SSE,
   except for MF 50 in 2015 and 2018.
This result reinforces the conclusion above.

\begin{table}[!htb]
\centering
\caption{$Alpha$ of indexes from 2015-2018}
\label{tbl:alpha}
\begin{threeparttable}
\begin{tabular}{c|c|c|c|c|c|c}
  \hline
  \multirow{2}{*}{}  & \multirow{2}{*}{Indexes} & \multicolumn{5}{c}{Number of constituents}\\
  \cline{3-7}
                         &                        & 50 & 100 & 150 & 180 & 380\\
  \hline
  \multirow{2}{*}{$2015$} & SSE                     &\textbf{-1.3965\%}	&2.2933\%	&4.7192\%   &-0.9256\%	&2.4680\%  \\
  \cline{2-7}
                                    & MF            &-1.6309\%	&\textbf{-0.7249\%}	&\textbf{-0.0524\%}	&\textbf{0.0142\%}	
                                                     &\textbf{0.6185\%}  \\
  \hline
  \multirow{2}{*}{$2016$} & SSE                     &0.6097\%	&-0.2126\%	&-0.1510\%	&0.2526\%	&-0.3369\% \\
  \cline{2-7}
                                    & MF            &\textbf{0.0569\%}	&\textbf{-0.0513\%}	&\textbf{-0.0464\%}	
                                                     &\textbf{0.0136\%}	&\textbf{0.0752\%} \\
  \hline
  \multirow{2}{*}{$2017$} & SSE                      &1.4407\%	&-0.3078\%	&-2.3713\%	&0.9818\%	&-0.4906\% \\
  \cline{2-7}
                                    & MF            &\textbf{-0.3905\%}	&\textbf{-0.2689\%}	&\textbf{-0.8270\%}	
                                                    &\textbf{-0.6286\%}	&\textbf{-0.4335\%} \\

  \hline
  \multirow{2}{*}{$2018$} & SSE                     &\textbf{0.5972\%}	&-0.5911\%	&-1.4430\%	&0.3667\%	&-0.7367\% \\
  \cline{2-7}
                                    & MF            &-0.8548\%	&\textbf{-0.3600\%}	&\textbf{0.0772\%}	&\textbf{-0.0696\%}	
                                                    &\textbf{0.0982\%} \\

  \hline
\end{tabular}
\end{threeparttable}
\end{table}

\emph{Beta} calculates the systematic risk of indexes in comparison to the benchmark market (SSE composite index).
We set the base line at 1.0;
  the closer \emph{Beta} is to 1.0,
  the closer is the risk of indexes from the market.
In Fig.~\ref{fig:4in2015}(c),
  the $Beta$ of MF 150 and MF 380 are closer to the base line than those of SSE 150 and SSE 380.
This result means,
  from the perspective of metric $Beta$,
  that MF 150 and MF 380
  perform more approximately to the stock market (SSE)
  than SSE 150 and SSE 380,
  respectively.
Meanwhile,
  from the perspective of systematic risk,
  the $Beta$ of MF indexes are all less than those of the SSE indexes,
  thus,
  the systematic risk of the MF index series is less than that of the SSE index series in 2015.
Combining the results of excess returns in Fig.~\ref{fig:4in2015}(b),
  MF indexes have lower excess returns and lower systematic risk,
  and SSE indexes have higher excess returns and higher systematic risk.
This result is a typical conclusion in investment theory.

\emph{Jensen's alpha}'s calculation considers both excess returns and systematic risk.
In  Fig.~\ref{fig:4in2015}(d),
  the result is similar to that of $Alpha$.
The $Jensen's Alpha$ of MF 100, MF 150, MF 180, and MF 380
  are closer to 0
  than those of SSE 100, SSE150, SSE 180, and SSE 380,
  respectively.
From the perspective of \emph{Jensen's alpha},
  the MF index series is more approximate to the stock market than the SSE index series.

\begin{figure}[!htbp]
   \begin{minipage}[t]{1.0\textwidth}
   \centering
    \subfigure[]
   {
        \includegraphics[width=0.4\textwidth]
          {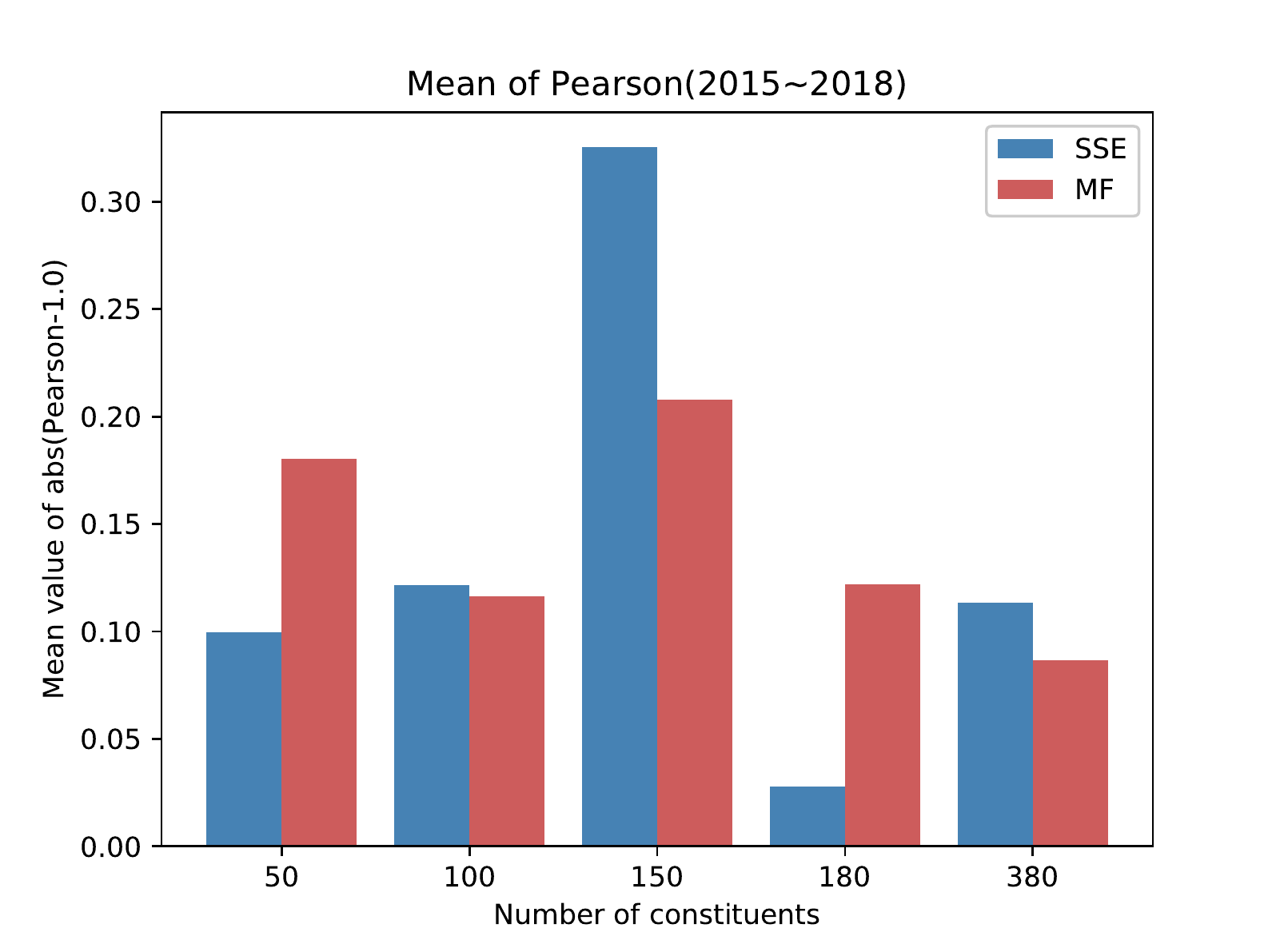}
   }
   \subfigure[]
   {
        \includegraphics[width=0.4\textwidth]
            {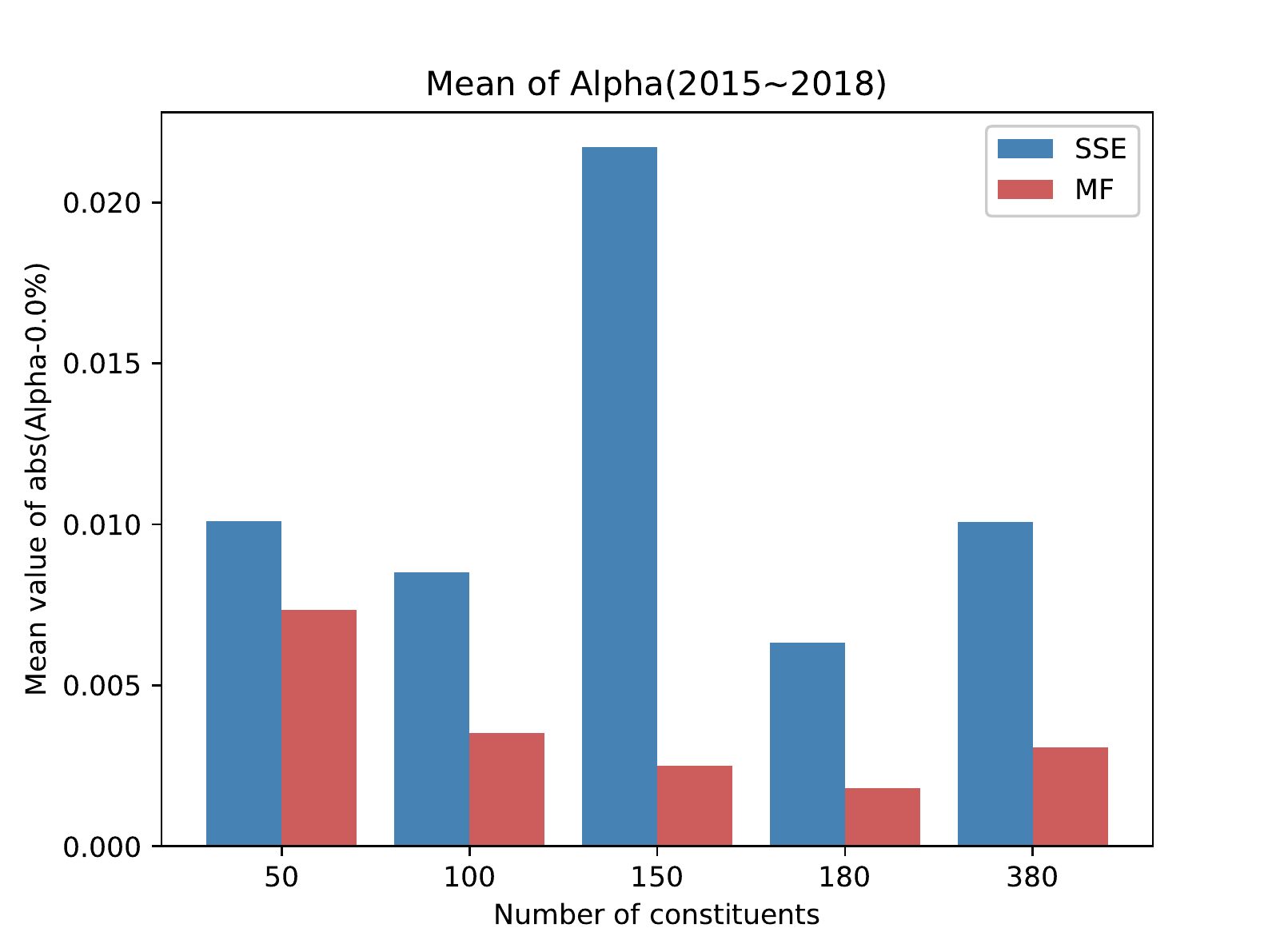}
   }
      \subfigure[]
   {
        \includegraphics[width=0.4\textwidth]
           {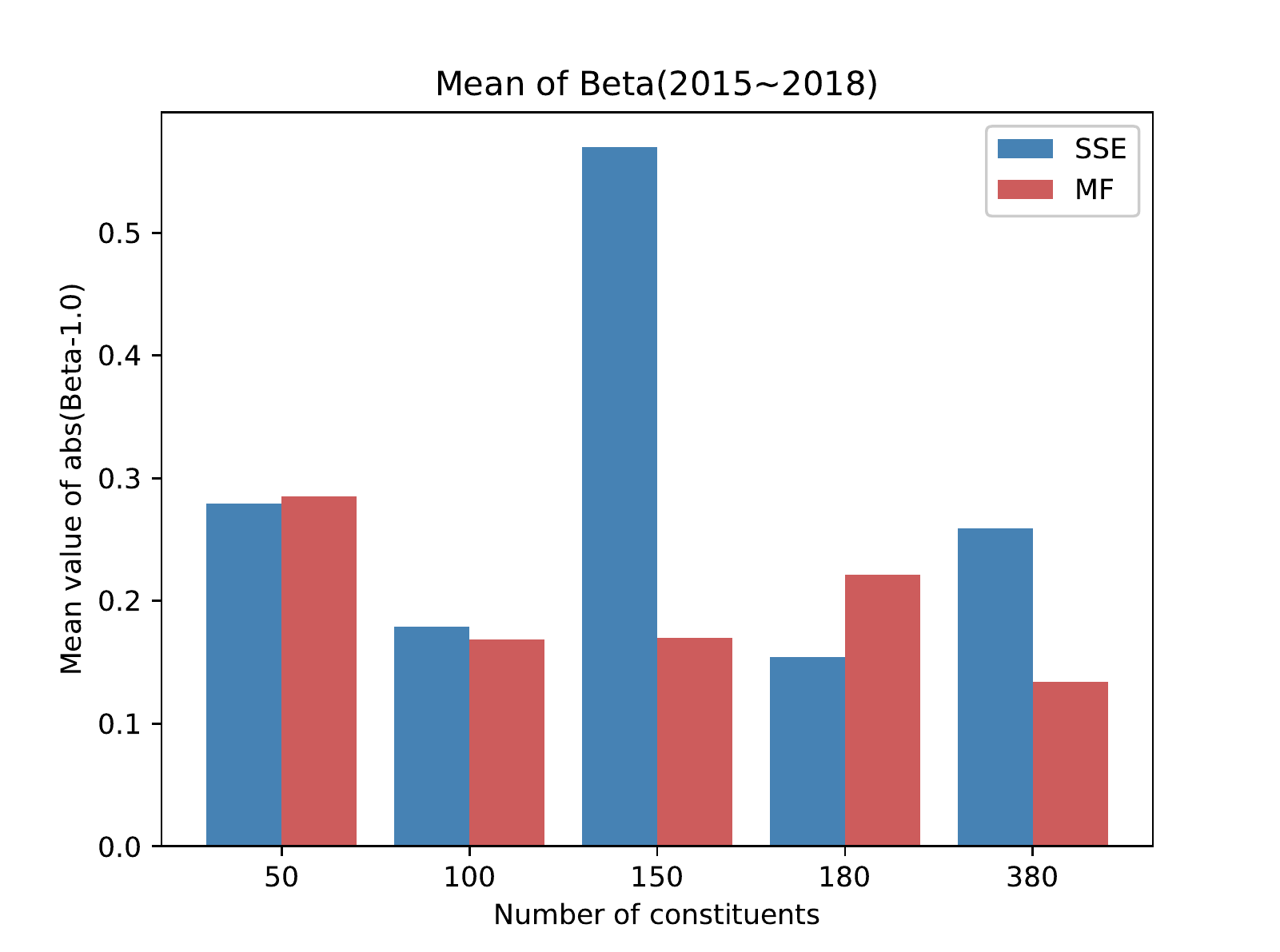}
   }
     \subfigure[]
   {
        \includegraphics[width=0.4\textwidth]
                  {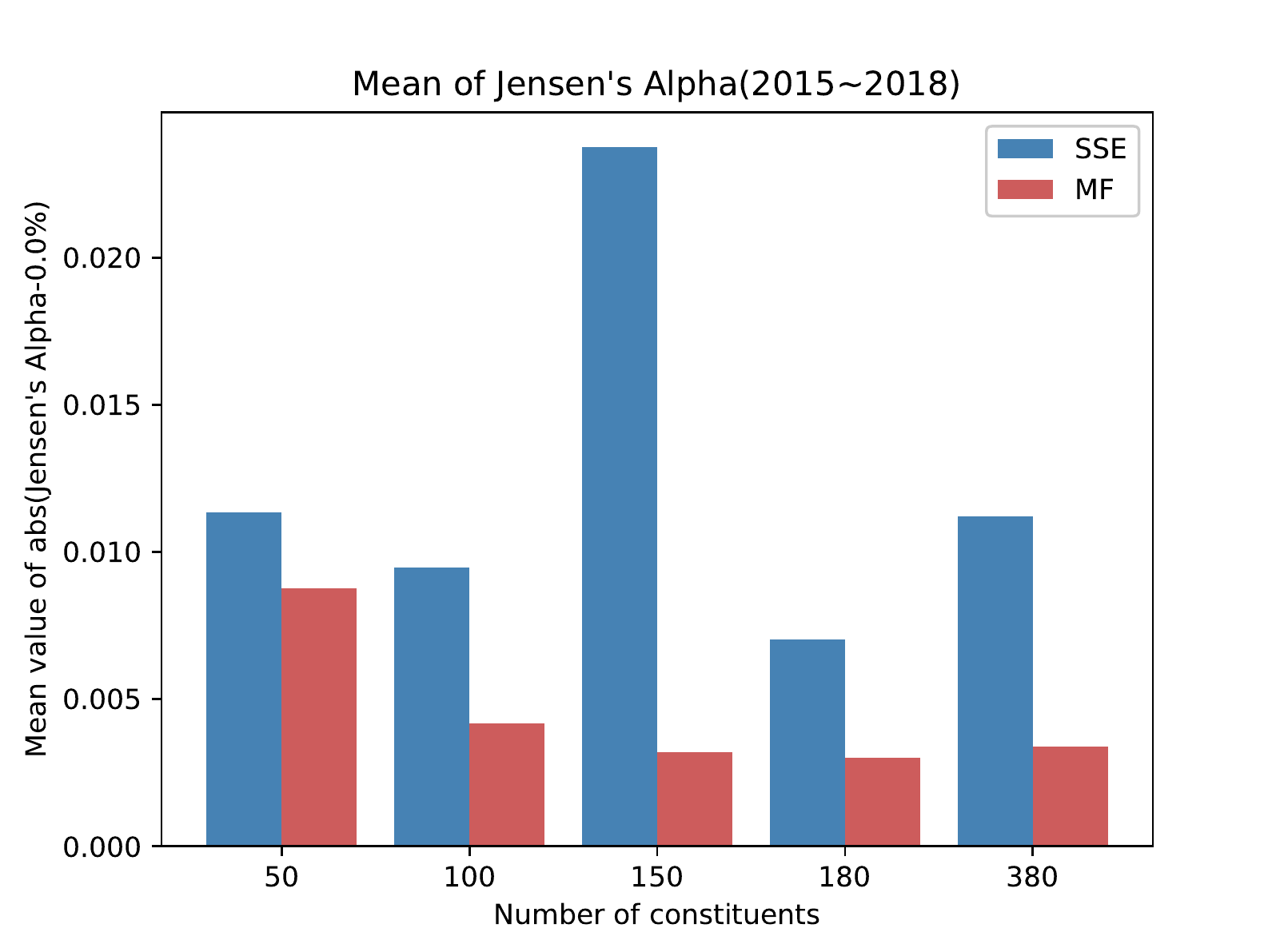}
   }
   \caption
   {
        Mean value of Pearson, Alpha, Beta and Jensen's Alpha from 2015 to 2018.
   }
   \label{fig:Mean}
   \end{minipage}
 \end{figure}

Summarizing all the results in Fig.~\ref{fig:4in2015},
   the MF index series is more approximate to the stock market than the SSE index series in 2015.
To make the conclusion more convincing,
   in Fig.~\ref{fig:Mean},
   we calculate the data from 2015 to 2018
   and show the mean value of the difference between the metric and the base line.
For example,
   in Fig.~\ref{fig:Mean}(a),
   to compare MF 50 and SSE 50 which is closer to the market,
   we calculate the mean value of the distance from the $Pearson$ of MF 50 to the base line as:
     \begin{equation*}
      \bar{D}_{Pearson} = \frac{\sum_{i=2015}^{2018} abs(Pearson_i^{MF 50}-1.0)}{4},
   \end{equation*}
   and the calculation of the $Pearson$ of SSE 50 is the same.
The results demonstrate that
   except for the $Pearson$ and $Beta$ of MF 50 and MF 180 (Fig.~\ref{fig:Mean}(a) and Fig.~\ref{fig:Mean}(c)),
   other metrics of the MF index series are all less than those of the SSE index series.
Finally,
   we conclude that the MF index series is more approximate to the stock market than the SSE index series.

\subsection{Stability}

\begin{figure}[!htbp]
   \centering
   \subfigure[]
   {
        \includegraphics[width=0.4\textwidth]
            {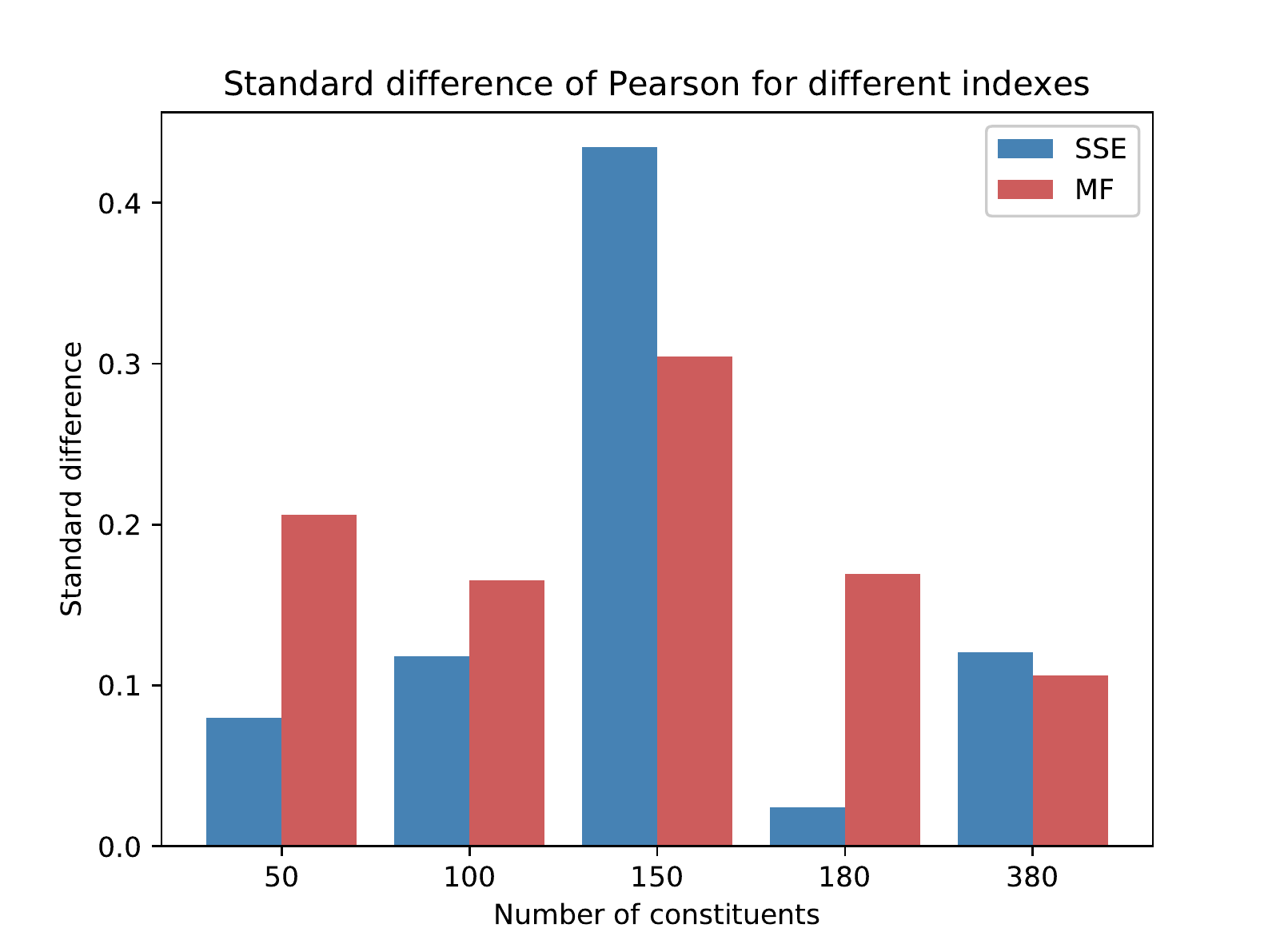}
   }
   \subfigure[]
   {
        \includegraphics[width=0.4\textwidth]
            {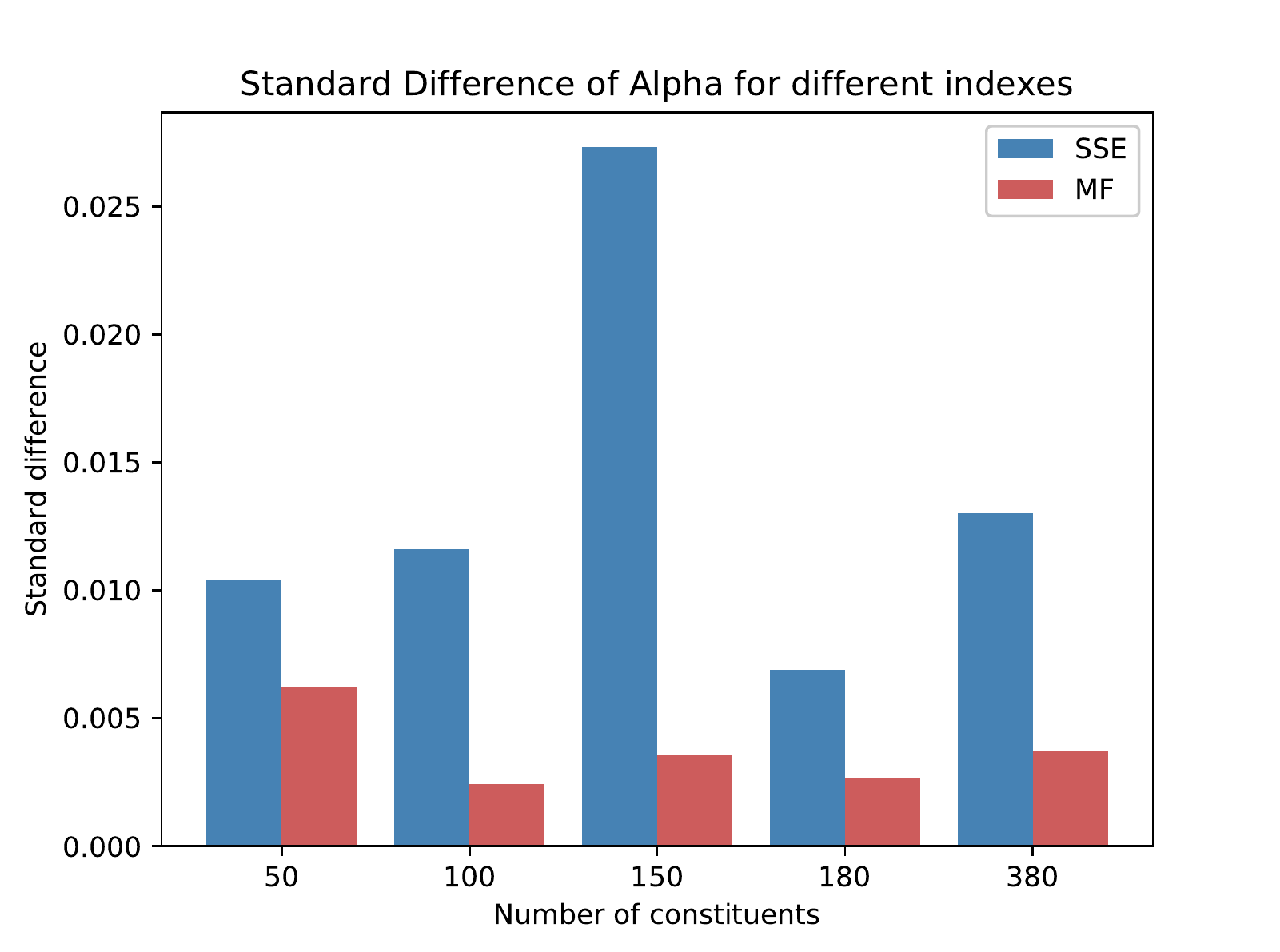}
   }
   \subfigure[]
   {
        \includegraphics[width=0.4\textwidth]
            {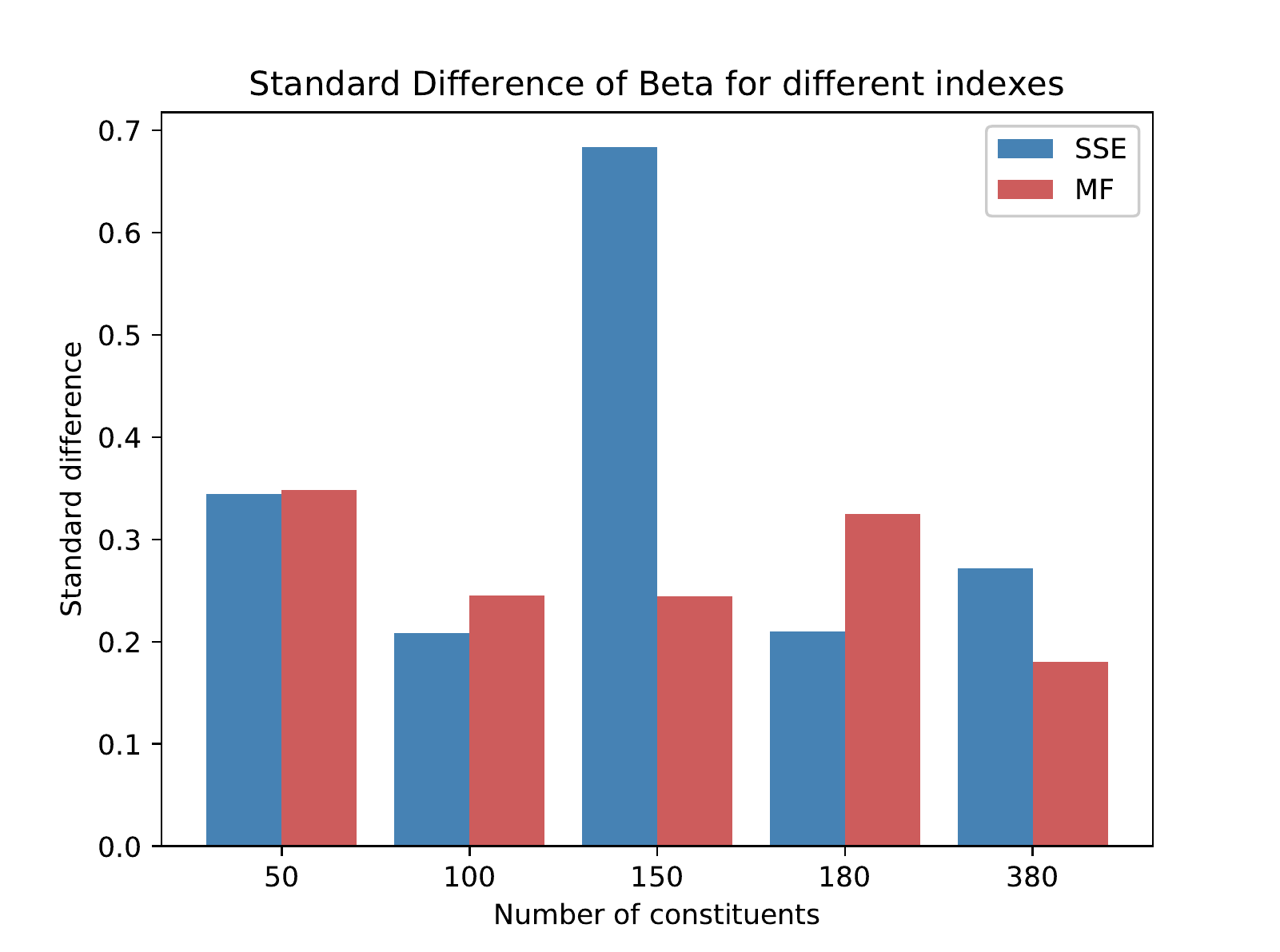}
   }
   \subfigure[]
   {
        \includegraphics[width=0.4\textwidth]
            {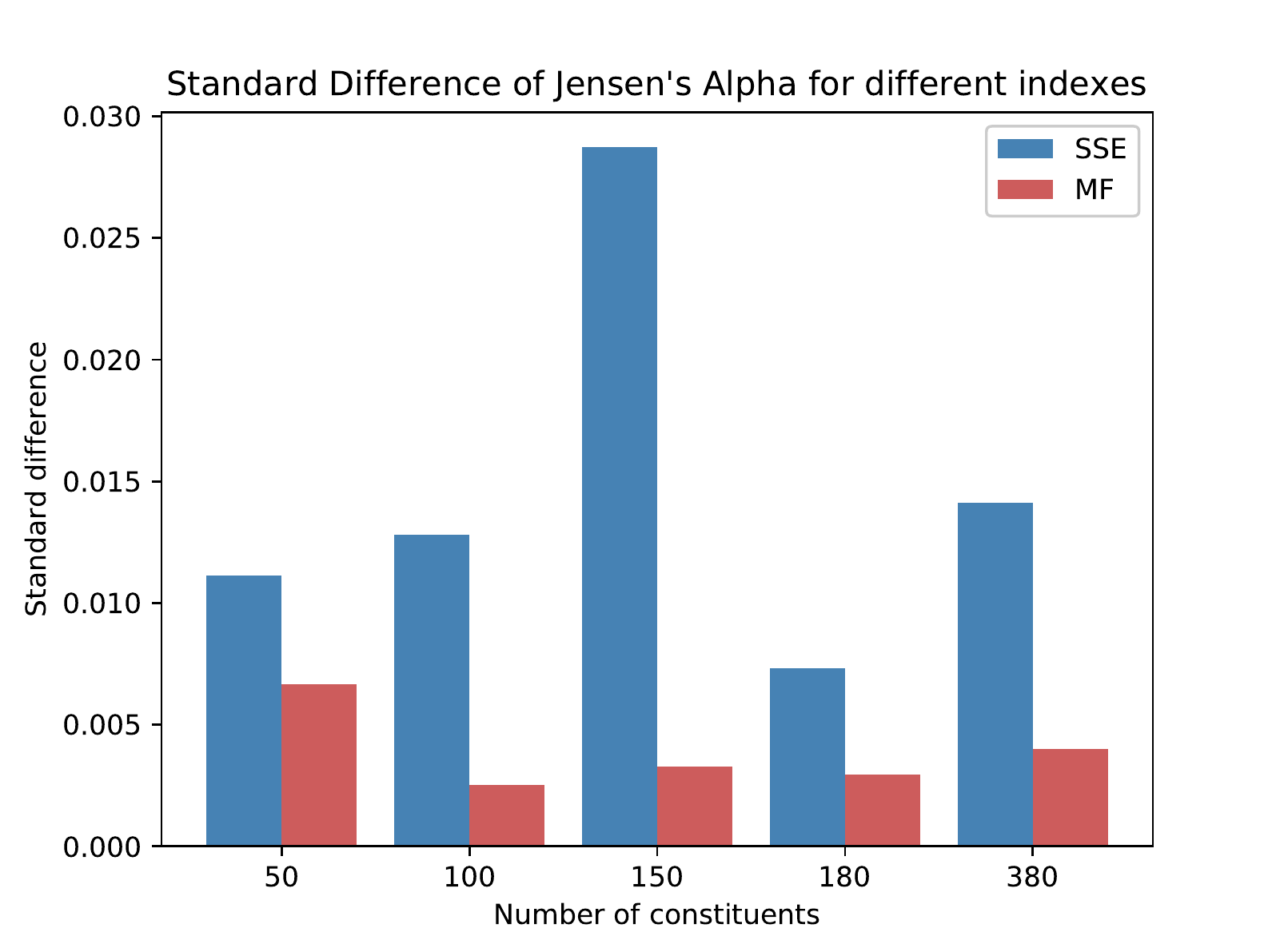}
   }
   \caption
   {
       The standard difference of four metrics from 2015-2018 for each index.
   }
   \label{fig:SDconstituent}
 \end{figure}

In this section,
  we discuss the stability of the MF indexes in two aspects.
On the one hand,
  for a metric of a single stock index,
  such as the $Pearson$ of MF 50,
  as MF 50 is designed to reflect the price activity of the market,
  the $Pearson$ of MF 50 is supposed to be always close to 1.0 in different year,
  which means the $Pearson$ of a good index is expected to be stable around 1.0.
We use the standard difference to describe the stability of the MF indexes,
  and we calculate the standard difference of the metrics,
  the closer the standard difference is to 0,
  the higher is the stability of the indexes.
In Fig.~\ref{fig:SDconstituent},
  we calculate the standard difference of metrics for each index in 4 years.
The results show that the stability of the MF index series is better than that of the SSE index series,
  especially in Fig.~\ref{fig:SDconstituent}(b) and (d),
  respectively.
In Fig.~\ref{fig:SDconstituent}(a) and Fig.~\ref{fig:SDconstituent}(c),
  the stability of MF 150 and MF 380 is better than that of SSE 150 and SSE 380,
   respectively.

\begin{figure}[!htbp]
   \centering
     \subfigure[]
   {
        \includegraphics[width=0.4\textwidth]
           {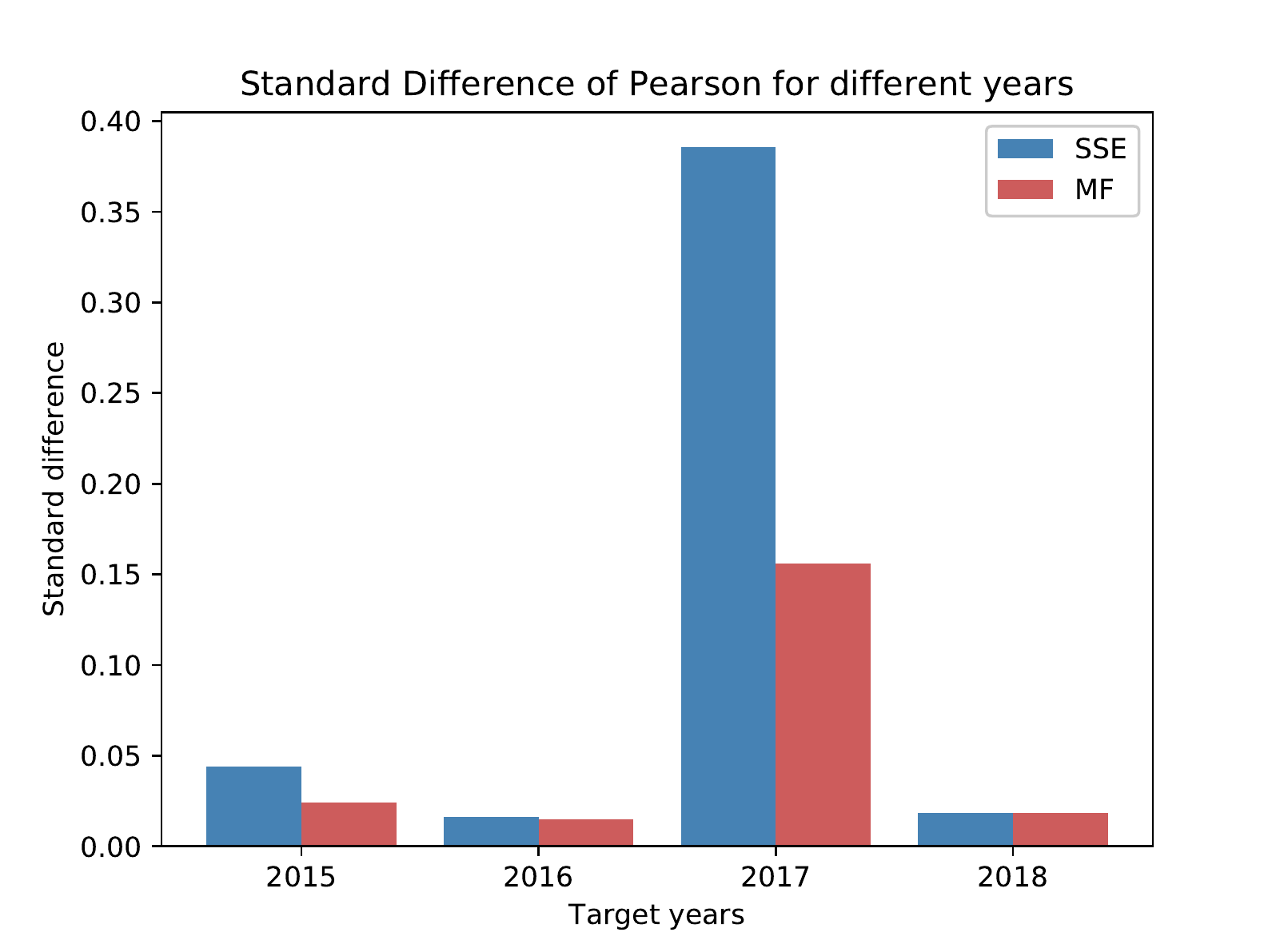}
   }
   \subfigure[]
   {
        \includegraphics[width=0.4\textwidth]
           {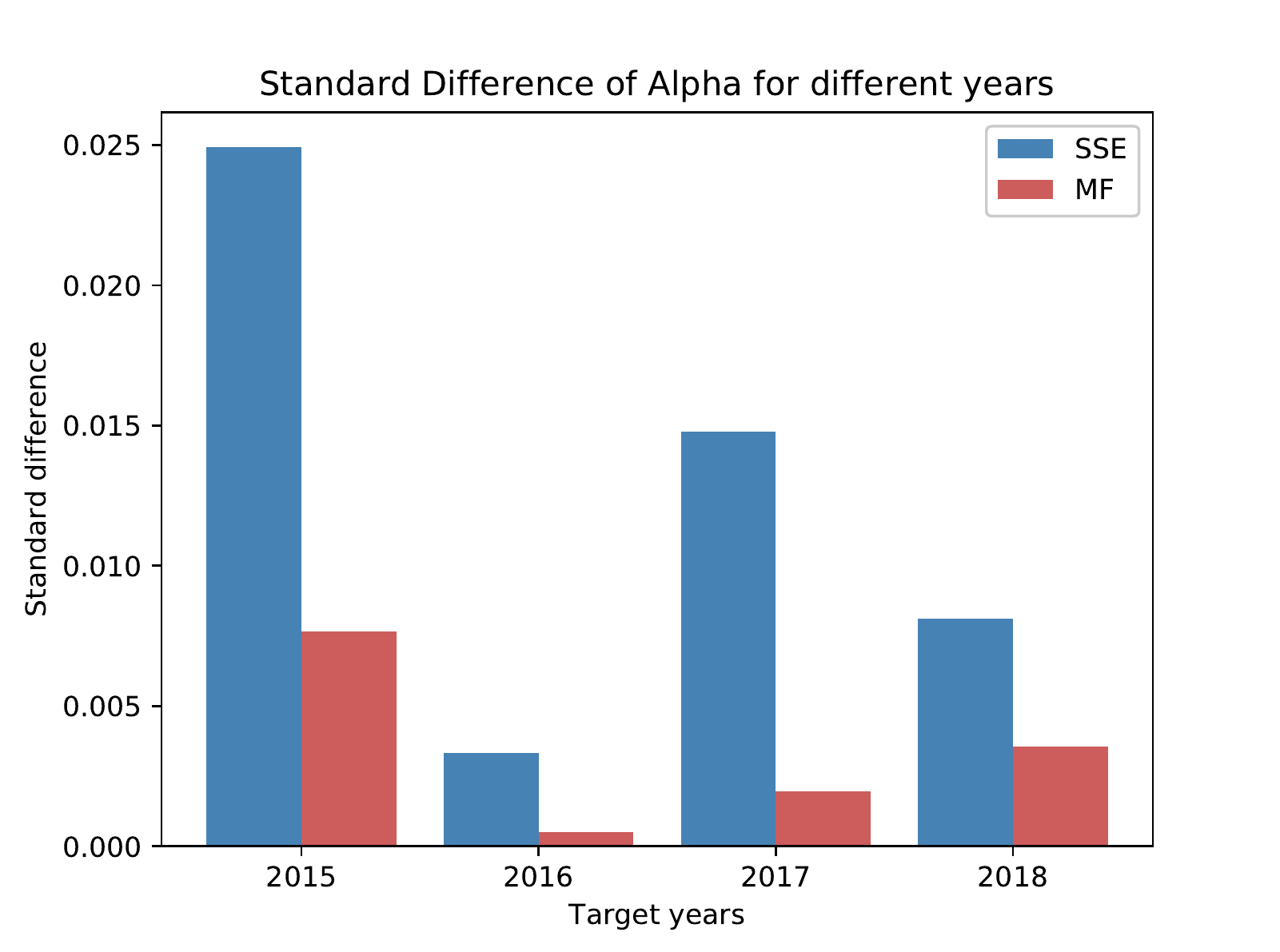}
   }
   \subfigure[]
   {
        \includegraphics[width=0.4\textwidth]
           {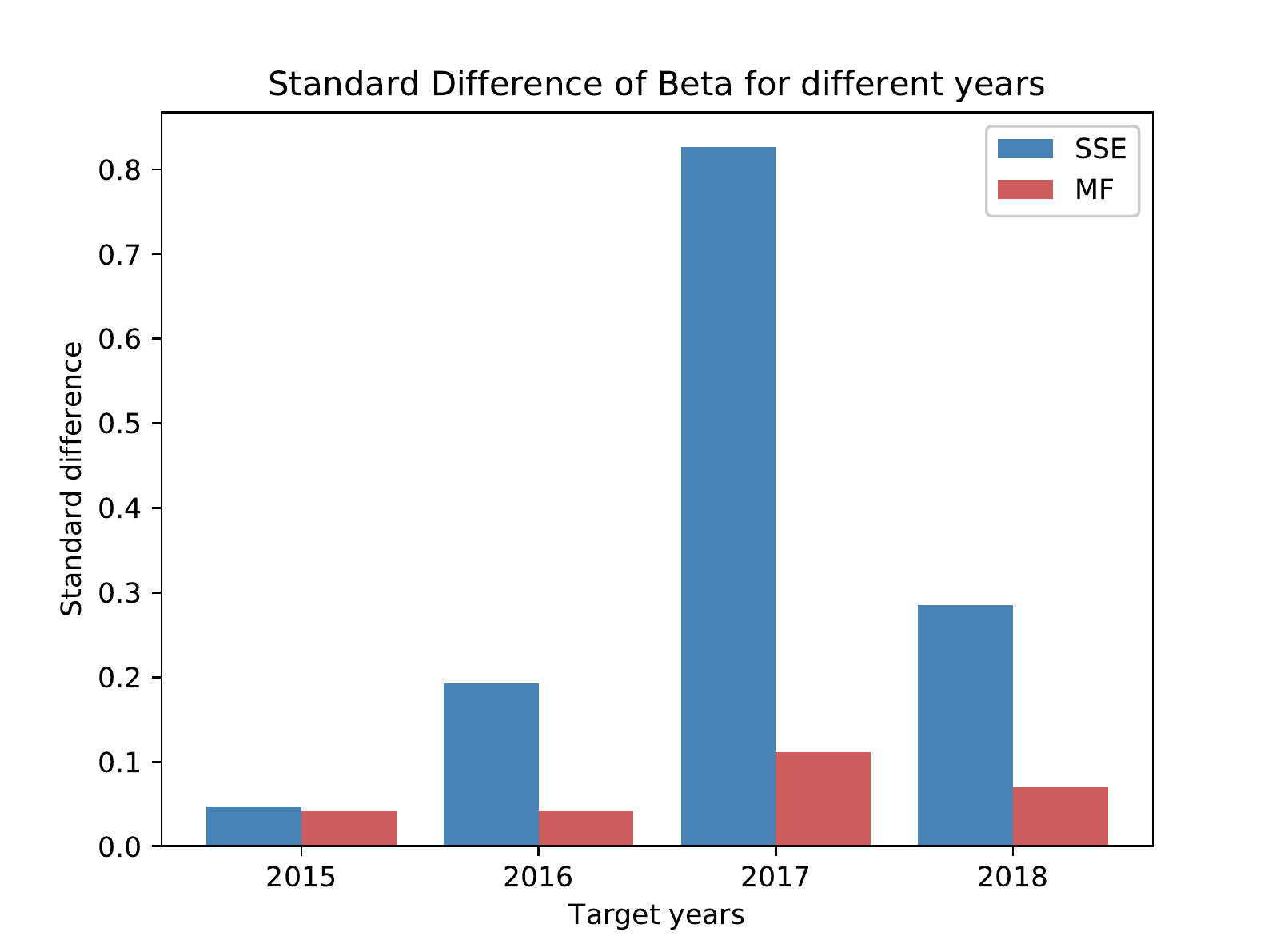}
   }
     \subfigure[]
   {
        \includegraphics[width=0.4\textwidth]
           {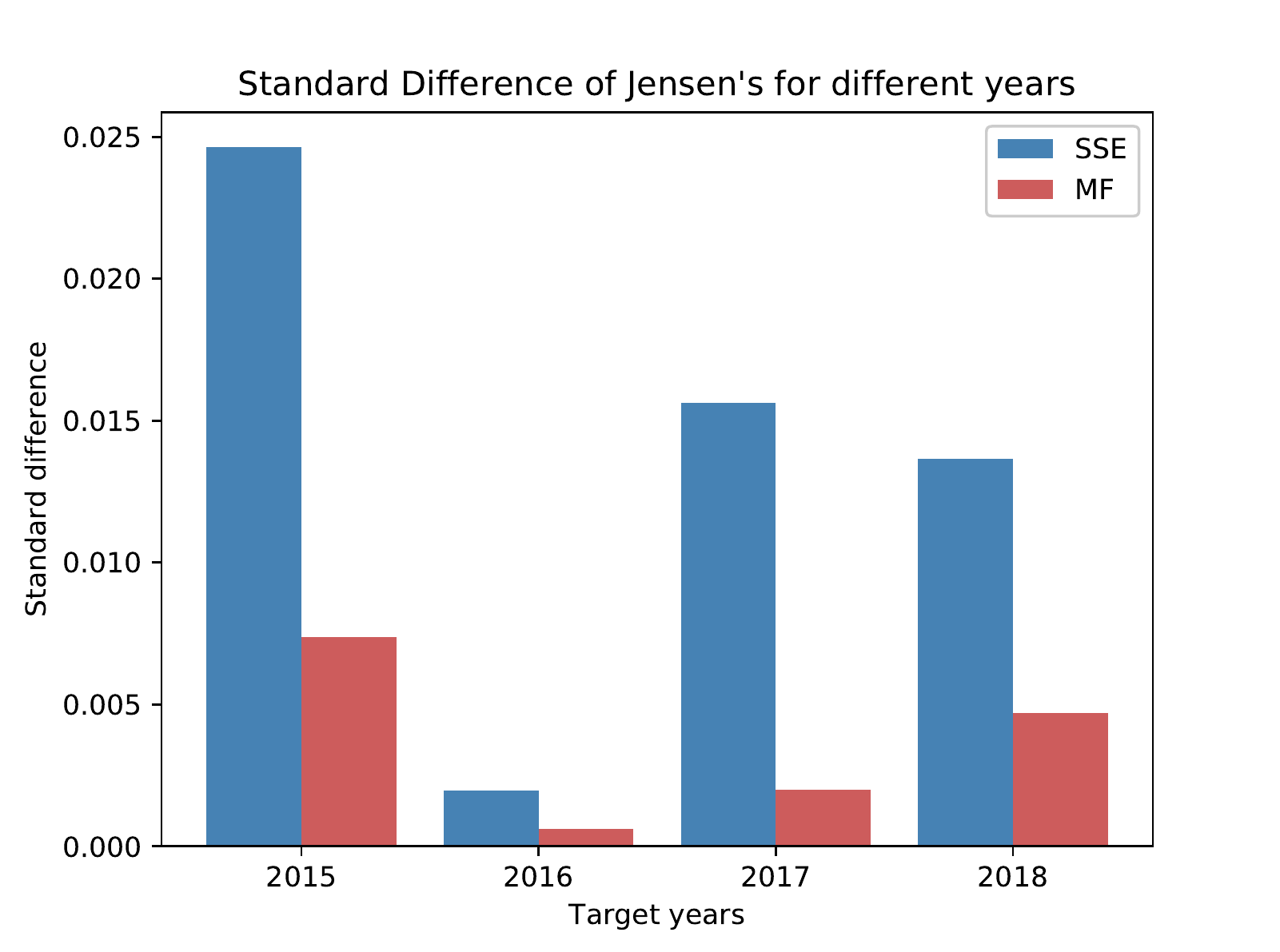}
   }
   \caption
   {
      The standard difference of MF index series(SSE index series)in each year.
   }
   \label{fig:SDyear}
 \end{figure}

On the other hand,
   for a metric of a stock index series,
   such as the $Pearson$ of the MF index series,
   when the constituent number $N$ increases,
   the $Pearson$ of MF index series is expected to gradually approach to 1.0.
A stable index series is expected to have stable $Pearson$ on the entire index series
   (MF 50, MF 100, MF 150, MF 180 and MF 380),
   then the MF index series can reflect the activity of the market stably.
In Fig.~\ref{fig:SDyear},
  we calculate the standard difference of the metrics for each index series per year.
The result shows that the MF index series is more stable than the SSE index series in every year.

An experimental result that can be found is that
  MF 50 performs worse than SSE 50 in almost all cases(Fig.~\ref{fig:4in2015}),
  but MF 100, MF 150, MF 180, and MF 380 do not.
The reason for this result may be that
  our high-dimensional simplification method approximates the market as the number of constituents increases,
  as in Fig.~\ref{fig:SSEVSMF2018}(c),
  but the SSE index series does not have such an approximation as the MF index series,
  as in Fig.~\ref{fig:SSEVSMF2018}(d).
Thus,
  MF indexes may perform worse than SSE indexes when they have few constituents;
  however,
  as the number of constituents increases,
  this result disappears.


\section{Conclusion}

In this paper,
   we introduce the construction and calculation method of the MF index series in detail.
By considering the stock dataset as a low-dimensional manifold embedded in a higher-dimensional Euclidean space,
   we construct its manifold structure and discrete LBO matrix.
After detecting the local maximum and minimum points on the eigenvectors of LBO,
   the stocks corresponding to these feature points are regarded as the constituents we selected from the complete stock space.
Moreover,
  this study details the weighting formula of the generation of the MF index and
  the index maintenance rules when the list of constituents changes or other non-trading factors are involved.
A complete calculation algorithm~\ref{alg:MFIndex} of the MF index series is given.
In the experiments conducted,
  we compare the MF index series and SSE index series using four metrics.
The results demonstrate that
   from the perspective of data approximation,
   our indexes are closer to the stock market (SSE composite index),
   and from the perspective of the risk premium,
   our indexes have higher stability and lower risk.

Finally,
  two points for future work are proposed.
First,
  in terms of the construction of the manifold structure,
  in this study,
  the descriptors of stocks only use high-dimensional closing prices.
They do not consider market cap, free-float, S/E, and other financial factors.
The advantage is that the relationship between different dimensions need not be considered
   when calculating the high-dimensional Euclidean distance.
However,
   a manifold structure based solely on the closing price is not the best.
Therefore,
   adding new factors,
   to establish a better manifold structure,
   is worth considering.
Second,
   with regard to the weighting formula of the final MF index,
   because this study considers approximating the MF index to the SSE composite index,
   it uses the same market cap weighting method as the SSE composite index without considering other financial factors such as the free-float.
However,
   in the existing stock market,
   some stock index weighting calculations consider more financial factors,
   such as sales,
   revenue,
   and earnings.
These weighting methods can also reflect the stock market sufficiently;
  therefore,
   modifying the weighting method is also worth considering.


\section*{Acknowledgements}
   This work is supported by the National Natural Science Foundation of China under Grant Nos.61872316,61932018.

\bibliography{mybibfile}

\end{document}